\documentclass[prl, twocolumn]{revtex4}
\usepackage{amsmath,amssymb,bm}
\usepackage{graphicx}
\usepackage{epstopdf}
\usepackage{latexsym}
\usepackage{subfigure}
\usepackage{color}
\usepackage{natbib}
\usepackage{braket}
\usepackage{hyperref}
\hypersetup{
  colorlinks,
  citecolor=magenta,
  linkcolor=blue,
  urlcolor=blue}
\usepackage{bbm}
\usepackage{xcolor}
\usepackage{enumitem}
\usepackage{float}
\usepackage{bibunits}

\begin{document}

\begin{bibunit}[apsrev]
\title{Spin-$S$ designer hamiltonians and the square lattice $S=1$ Haldane nematic}

\author{Nisheeta Desai}
\author{Ribhu K. Kaul}
\affiliation{Department of Physics \& Astronomy, University of Kentucky, Lexington, KY 40506-0055}

\begin{abstract}
We introduce a strategy to write down lattice models of spin rotational symmetric
Hamiltonians with arbitrary spin-$S$ that are Marshall positive and can be simulated
efficiently using world line Monte Carlo methods. As an
application of our approach we consider a square lattice $S=1$ model
for which we
design a $3\times 3$ - spin plaquette interaction. By numerical simulations we
establish that our model realizes 
a novel ``Haldane nematic'' phase that breaks lattice rotational
symmetry by the spontaneous formation of Haldane chains,
while preserving spin rotations, time reversal and lattice
translations. By supplementing our model with a two-spin
Heisenberg interaction, we present a study of the transition between N\'eel and Haldane
nematic phase, which we find to be of first order.
\end{abstract}
\date{\today}
\maketitle

{\em Introduction:} 
The relationship between lattice spin models and their long distance
descriptions by quantum field
theories is a central topic in theoretical condensed matter physics~\cite{sachdev1999:qpt,fradkin2013:book}.
Pioneering work on the ground state of spin chains found a striking role is played 
by the size of the quantum spin~\cite{haldane1988:berry,affleck1989:lh}:
while half integer spins generically realize a gapless critical phase,
integer spin chains realize a topological ``Haldane phase''. In the
field theoretic understanding, the value
of the microscopic value of the spin enters as a co-efficient of a
topological term that has a dramatic effect on the spin chain phase diagram. Given
this profound result
in one dimension, it is natural to ask how the value of the
spin-$S$ affects the phase diagrams of two
dimensional quantum spin systems?

For one dimensional systems, progress in our understanding is largely due to
the availability of specialized
analytic~\cite{sutherland2004:bm,giamarchi2004:one} and numerical
methods~\cite{white1992:dmrg}. These methods 
cannot be extended as effectively to two dimensions, where consequently
much less is known despite intense research. 
The most reliable unbiased method to study field theory and quantum criticality in two dimensions are limited to models that do not suffer from the sign problem
of quantum Monte Carlo~\cite{kaul2013:qmc}. Although the sign-free
condition is very restrictive,
given their unique ability to provide unbiased insight it is of great interest to build a
repertoire of sign-free spin models for arbitrary spin-$S$, as has
been achieved for $S=1/2$~\cite{kaul2014:design}.

\begin{figure}[t]
 \includegraphics[trim={0 11cm 5cm 0},clip, width=5in]{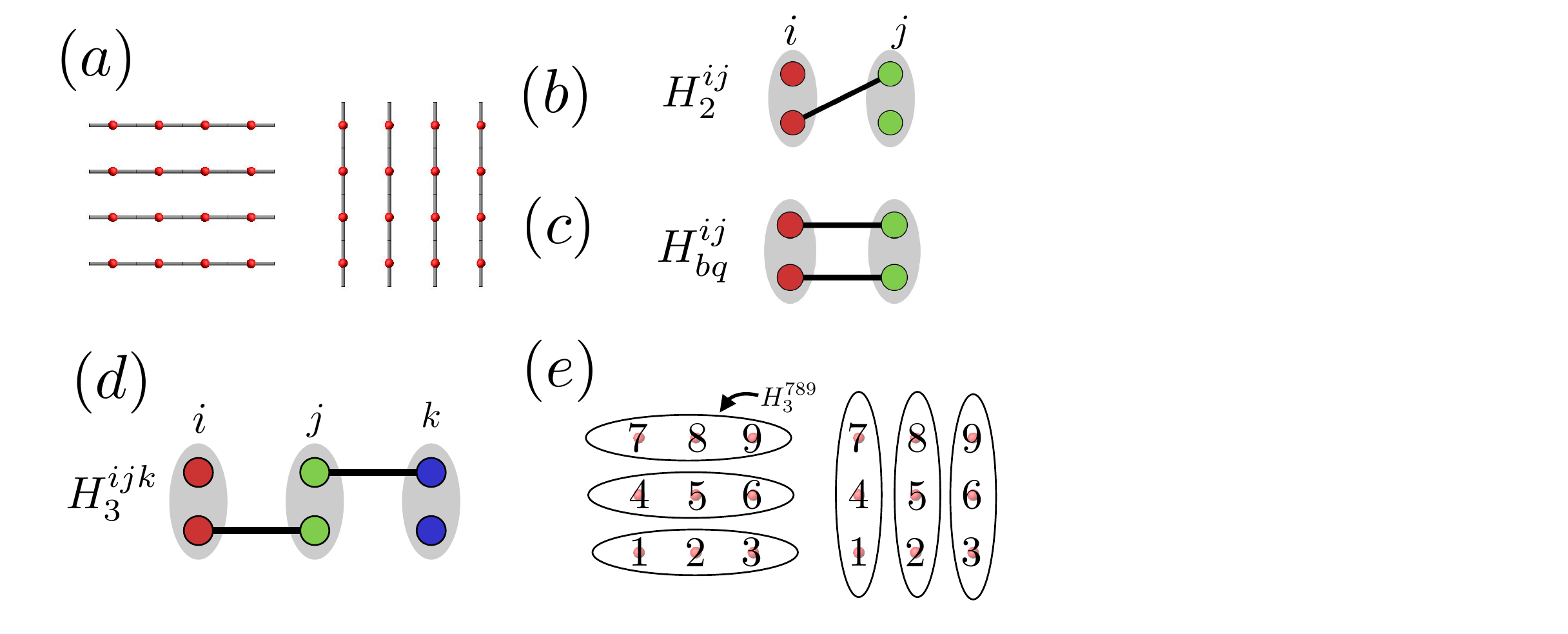} 
\caption{\label{fig:intro} (a) Two-fold degenerate
  ground states in the ``Haldane nematic'' phase for square lattice
  $S=1$ spins (red spheres). The strong bonds
  indicate the spontaneous formation of gapped Haldane chain that 
  breaks lattice rotational symmetry while preserving 
  translations. (b,c,d) Representative mini-spin interaction diagrams that
appear in the  (b) two-site $H^{ij}_2$ Heisenberg, Eq.(\ref{eq:heis}),
(c) the
$H^{ij}_{bq}$ biquadratic, and (c) three-site $H^{ij}_3,$
Eq.(\ref{eq:3spin}) interactions.
The two mini-spins corresponding to an
on-site $S=1$ are collected in a grey bubble. (e) The $H^p_{3\times
  3}$ interaction acts on the elementary $3\times 3$
plaquette indexed by $p$. It is constructed out of sum of two terms,
each of which is a product of three $H^{ijk}_3$
terms. To preserve square lattice symmetry both the orientations
that are shown are included in Eq.~(\ref{eq:q33}).}
\end{figure}

In this Letter we develop a systematic method to write down a large
family of sign-free bipartite spin models with arbitrary spin-$S$ and multi-spin
interactions that have the
Heisenberg rotational symmetry. These
new models open the door to study a variety of new phases and phase
transitions, many of which are of great interest to the
community. 
As a first application of our method we design a
square lattice $S=1$ interaction that realizes a long anticipated ``Haldane
nematic'' (HN)
phase~\cite{affleck1988:aklt,read1990:vbs}.  In this phase the spin
system breaks lattice rotation
symmetry but preserves lattice translations due to the spontaneous formation of
Haldane chains either in the $x$ or $y$ direction with an associated two-fold ground
state degeneracy, Fig.~\ref{fig:intro}(a). Motivated in part by
the Iron superconductors the HN phase has been under intense
study recently~(see e.g. \cite{chen2018:s1,jiang2009:s1,gong2017:s1,bilbao2015:nem,corboz}). An
influential work~\cite{wang2015:s1} found an exactly solvable model which realizes the HN as
a ground state and provided field theoretic arguments for an exotic
continuous phase transition to a N\'eel state described by the O(4)
$\sigma$-model at $\Theta=\pi$. We establish unambiguously the existence of the
HN phase in our new sign free model and
provide the first unbiased numerical study of the phase
transition from the HN to the N\'eel state. We find clear evidence that the
transition is first order and  discuss
the implications of this finding for the field theoretic scenario. 

{\em Designer Models:}
While it is well known that the bipartite Heisenberg model is
Marshall positive for arbitrary spin-$S$, what are the most general multi-site spin-$S$ Hamiltonian operators that are
sign positive? This question has been difficult to address
previously because it appears daunting directly in the
language of spin-$S$ operators. Following previous
work~\cite{todo2001:highs, kawashima1994:spinS,kawashima2004:review} we take a
different route -- we rewrite the spin-$S$ on each of the $N_s$ lattice sites as $2S$ spin-1/2
``mini-spins'', 
\begin{equation}
\label{eq:Ssig}
{\bf S}_i = \sum_a {\bf s}^a_i.
\end{equation}
We note here that the ${\bf s}^a_i$ have both a lattice index $i$ ($1\leq i
\leq N_s$) and a
mini-spin index $a$ ($1\leq a
\leq 2S$), giving a total of $2S N_s$ mini-spins.
 To faithfully
simulate the original problem, we have to include a projection
operator, ${\cal P}=\prod_i {\cal P}_i$, where ${\cal P}_i$ projects
out the
spin-$S$ from the ${\bf s}^a_i$ basis, $ Z= {\rm Tr}_{\bf S} \left [
  e^{-\beta H( {\bf S})}\right ]= {\rm Tr}_{\bf s}
\left [ e^{-\beta H( {\bf s})} {\cal P}\right ]$. Since ${\cal P}$
is itself sign-problem free, in the world-line approach, any model which is
sign-free in the ${\bf s}^a_i$ basis gives us a sign-free spin-$S$ model!

In this manuscript we illustrate our idea using $S=1$ spins on the square
lattice, but our results can be straightforwardly extended to any
bipartite lattice with arbitrary spin-$S$. 
Consider first in the ${\bf s}$ language the $S=1$
Heisenberg model, 
\begin{equation}
\label{eq:heis}
H^{ij}_{2} =  {\bf S}_i \cdot {\bf S}_j -1 = -\sum_{a,b} (\frac{1}{4}- {\bf s}^a_{i} \cdot {\bf s}_{j}^b )
\end{equation}
Diagramatically we can represent each $\frac{1}{4}-{\bf s}^a_i\cdot
{\bf s}^b_j$ term in the sum in the last
expression as an ``${\bf s}$-bond'' between
mini-spins $a$ and $b$ on
the two sites $i$ and $j$. A representative such term is illustrated for $S=1$
with two mini-spins per site in Fig.~\ref{fig:intro}(b) (there are
three other such diagrams corresponding to the sum on $a,b$). 
Likewise, it is easy to see that the interaction with two ${\bf s}$-bonds between $i$ and $j$
corresponds to the sign free region of the biquadratic
interaction,  Fig.~\ref{fig:intro}(c) ~\cite{harada2001:bq,spsdes2019:supmat}.
From these
examples, we make our central observation -- {\em it is much easier to write down a sign free
model in the ${\bf s}$ language than directly in the spin-$S$
basis}.   As a non-trivial
example consider interactions between three $S=1$ spins in a row. In
the ${\bf s}$-bond language the most natural interaction is with a single
bond between each pair of neighbors without allowing them to touch on
the middle site, Fig.~\ref{fig:intro}(d). Working
backwards we then find this new sign-free interaction in terms of the
spin-1 operators is,
\begin{eqnarray}
\label{eq:3spin}
H^{ijk}_3 &=&-{\bf S}_i \cdot {\bf S}_j {\bf S}_j \cdot {\bf S}_k -{\bf S}_k
\cdot {\bf S}_j {\bf S}_j \cdot {\bf S}_i\\
&+& {\bf S}_i \cdot {\bf S}_j + {\bf S}_i \cdot {\bf S}_k + {\bf S}_j \cdot {\bf S}_k-1
\end{eqnarray}
For $S=1$ models the three-site interaction and its physical significance  has been
discussed recently~\cite{michaud2013:s1,chepiga2016:s1}. Here we
discover that in order to study such terms in a sign free way we have
to include two spin terms to balance the signs. Intuitively, the three spin
interaction in Fig.~\ref{fig:intro}(d) is reminiscent of the famous AKLT
construction~\cite{affleck1988:aklt} and so we can expect it to force our system into a
Haldane like phase; we confirm this below. Using the three-site
interaction $H^{ijk}_3 $, we introduce a model interaction
we will study in detail below. Following the idea of the J-Q model~\cite{sandvik2007:deconf} we
construct a $3\times 3$ plaquette interaction from $H^{ijk}_3$,
\begin{equation} 
\label{eq:q33}
H^p_{3\times 3} = H^{123}_3 H^{456}_3 H^{789}_3 + H^{147}_3 H^{258}_3 H^{369}_3 ,
\end{equation} 
The indexing of the sites  in the plaquette by numbers 1-9 is shown in
Fig.~\ref{fig:intro}(e). The two terms are included to preserve square
lattice symmetry \footnote{$3\times 1$ and $3\times 2$ plaquettes interactions
can also be considered, but they are found to be unsuitable for the
application, i.e. they are insufficient to destroy N'eel
order~\cite{spsdes2019:supmat}. A $S=1$ version of the $Q_3$ term ~\cite{sandvik-kawashima} is also found unable to destabilize N'eel order. }. 

We emphasize that in addition to the advantage of leading us to new non-trivial sign
free interactions, the mini-spin representation also offers us a
simple way to construct efficient loop update algorithms for complex
interactions such as Eq.~(\ref{eq:q33}), since we can
update the ${\bf s}$ interactions using the standard deterministic
algorithm using for e.g. the stochastic series expansion~\cite{sandvik2010:vietri}. The update of the symmetrization operator is
straightforward using the directed loop algorithm ~\cite{spsdes2019:supmat,directed-loops}.  Clearly this
program of designing sign-free interactions in terms of the ${\bf
  s}$-bond diagrammatic
representation and then into the spin operators can be extended
systematically to any
value of spin-$S$ and to a wide range of multi-spin
interactions. Rather than elaborate on this here, we now turn to an
application.

\begin{figure}[t]
 \includegraphics[trim={0 0 0 0},clip, width=3in]{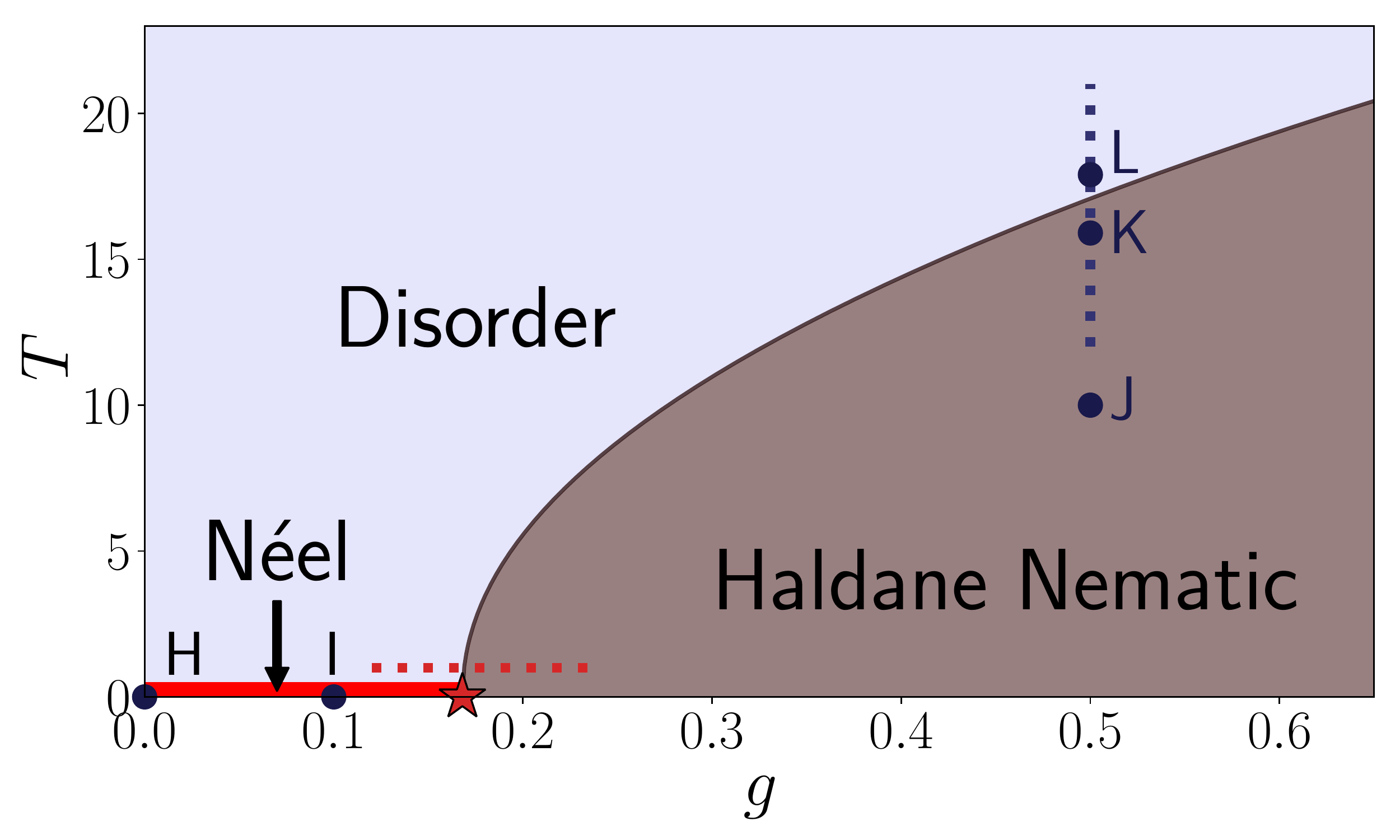} 
\caption{\label{fig:pd} Phase diagram of the model
  Eq.~(\ref{eq:Hg}) in the $g$-$T$ plane. As we establish by extensive numerical simulations, at $g\approx 0.17$ there is
  a first order quantum phase transition from N\'eel to the Haldane nematic (HN)
  phase. The solid line is a guide to the eye
of the  phase transition between HN and a simple disordered
  phase. The solid line is determined as a fit to the location of the
  transition at a few different
  $g$ \cite{spsdes2019:supmat} by
  detailed numerical study (as in Figs.~\ref{fig:ising},\ref{fig:first}). The transition is found to be continuous at high-$T$ and first order at
  low-$T$ (close to the quantum transition), see respectively Fig.~\ref{fig:ising}
  (corresponding to the vertical dashed line at $g=0.5$)
  \& Fig.~\ref{fig:first} (the horizontal dashed line at $T=1$). The
  change from first order to continuous Ising is known to take place at a
  tricritical Ising point - locating the tricritical point
  in our phase diagram is beyond the scope of this work, its location does not affect our conclusions.  The $(g,T)$ values for the points labeled in the
  phase diagram and presented in Fig.~\ref{fig:op} are {\bf H}:(0,0) {\bf I}:(0.1,0) {\bf J}:(0.5,10) {\bf K}:(0.5,15.9) {\bf L}:(0.5,17.9) }
\end{figure}

{\em Haldane Nematic:} We
consider square lattice $S=1$ antiferromagnets, which have been argued to host an exotic ``Haldane
nematic'' (HN) state in their phase diagrams. Our goal here is to establish
that the sign-free model, Eq.~\ref{eq:q33} realizes this novel phase and carry our
unbiased studies of the phase
transitions of the destruction of HN order.

The model we study is,
\begin{equation}
\label{eq:Hg}
H = J\sum_{\langle ij \rangle} H^{ij}_2 +Q_{3\times 3} \sum_{p} H^{p}_{3\times 3}.
\end{equation}
The first term is the usual square lattice $S=1$ Heisenberg model. The
second term is our new designer interaction with a sum on $p$, which
runs over the elementary $3\times 3$ plaquettes on the square
lattice.  We study the 
phase diagram as a function of $g\equiv Q_{3\times 3}/J $ and the temperature
$T=1/\beta$. We work in units in which $J^2+Q^2_{3\times 3}=1$.
The phase diagram inferred from our simulations is shown
in Fig.~\ref{fig:pd}. At $(g,T)= (0,0)$ (labelled as {\bf H}) our
model is the nearest neighbor
$S=1$ Heisenberg model which is N\'eel ordered~\cite{singh1990:s1}. We use the
conventional order parameter $\langle m^2 \rangle$ with 
$m = \sum_{\bf r} e^{i (\pi,\pi)\cdot {\bf r}} S^z_{\bf r}/N_s$ 
to diagnose long range magnetic
order.  From the finite size scaling of $\langle m^2 \rangle$ we
observe that the N\'eel order
weakens as $g$ is increased ({\bf I}). At $T=0$ the N\'eel
order is stable until we reach a coupling $g\approx 0.17$ at which
N\'eel order is destroyed. As is well known, the N\'eel
order cannot survive finite-$T$ Mermin-Wagner fluctuations in two dimensions.

\begin{figure}[t]
 \includegraphics[trim={0 0 0 0},clip, width=3.5in]{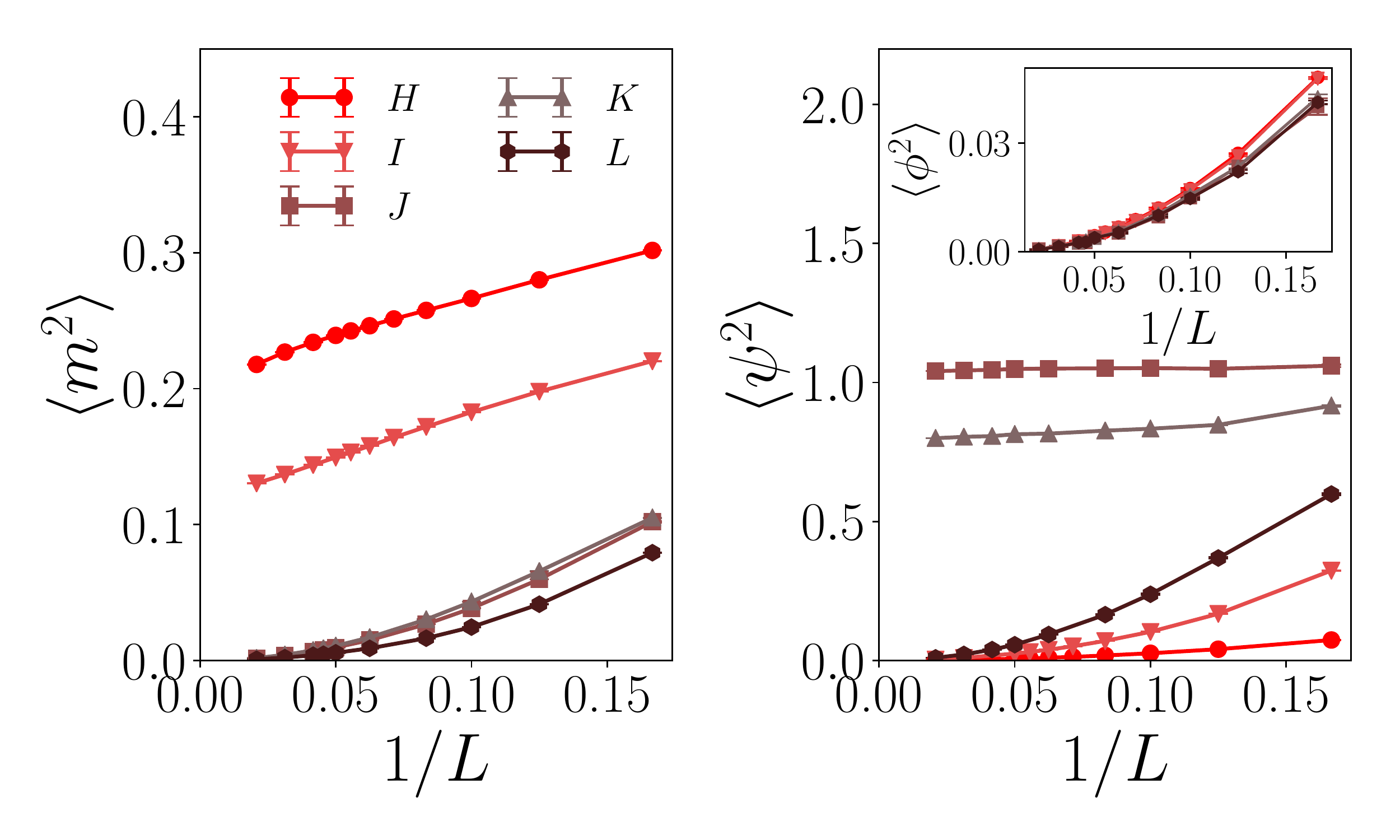} 
\caption{\label{fig:op} Extrapolations of the order parameters as a
  function of $1/L$ for various points labeled in the $g$-$T$ phase
  diagram shown in Fig.~\ref{fig:pd}. The left panel shows the N\'eel
order parameter, the right panel shows the order parameter for the
Haldane nematic. The inset on the upper right shows the
conventional ``dimerized'' $(\pi,0)$ VBS
order, $\langle \phi^2\rangle$ that breaks translations as well as
rotations, which is found to vanish in the model under study here.}
\end{figure}

\begin{figure}[t!]
 \includegraphics[trim={0 0 0 0},clip, width=3.5in]{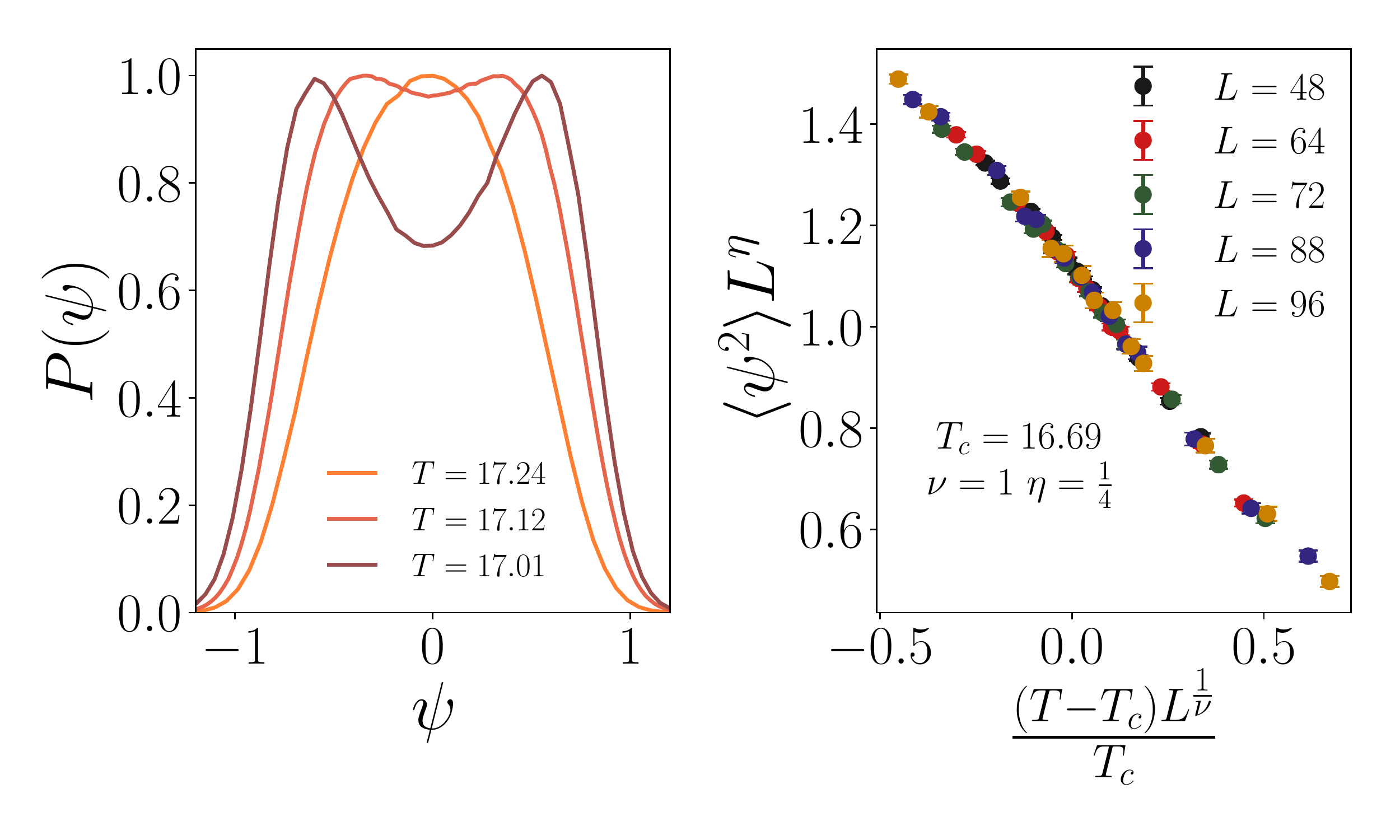} 
\caption{\label{fig:ising} Behavior of the HN order parameter
  at the thermal transition at $g=0.5$,  along the vertical dashed line in Fig.~\ref{fig:pd}. The left panel shows the
  histograms of the order parameter $\psi$ as $T$ is lowered
showing the emergence of two symmetry related Ising peaks, and no evidence
for first order behavior or phase co-existence. The right
  panel shows the collapse of the HN order parameter with two
  dimensional Ising critical exponents, providing further evidence for
a two-fold symmetry breaking in the ground state, consistent with Fig.~\ref{fig:intro}(a).}
\end{figure}
We now present extensive numerical evidence
that at $T=0$ for $g\geq 0.17$ the system transitions into the ``Haldane nematic''
phase (Fig.~\ref{fig:intro}(a)).  We first rule out
a conventional VBS pattern where pairs of $S=1$ dimerize into a
columnar pattern~\cite{wildeboer2018:s1}, which can be studied by finite size scaling of $\langle \phi^2 \rangle$ with $\phi
= \sum_{\bf r}e^{i (\pi,0)\cdot {\bf r}}B_x({\bf r})/N_s$ [with the
bond operator $B_{i}({\bf r})\equiv J{\bf S}_{\bf r}\cdot
{\bf S}_{{\bf r}+{\bf e}_i}$]. As shown 
in the inset of Fig.~\ref{fig:op} $\langle \phi^2\rangle$ scales to 
zero in the thermodynamic limit indicating that in all parts of the phase diagram under study the
conventional VBS order is absent. We use an order parameter~\cite{okubo2015:sun}
$\langle \psi^2\rangle$ that is sensitive to breaking of rotational
symmetry without picking up signals of translational symmetry
breaking. $\psi = \sum_{\bf r}(B_x({\bf
  r})-B_y({\bf r}))/N_s$. Clearly a condensation of $\psi$ indicates the
breaking of lattice rotational symmetry. As shown in Fig.~\ref{fig:op}
{\bf K} and {\bf J} clearly have long range HN order, whereas at the other
points they are absent either because of N\'eel order ({\bf H} and
{\bf I}) or thermal disorder ({\bf L}).

We now turn to a study of the phase transition at which HN order is
destroyed. We begin by simulating the model at $g=0.5$ and tuning
$T$ along the vertical dashed line in Fig.~\ref{fig:pd}. From Fig~\ref{fig:op}, as we move from {\bf L} (no HN order) to
{\bf K} (HN order) to {\bf J} (stronger HN order) we have clear
evidence for a phase transition. If the
pattern of symmetry breaking is of the form
Fig.~\ref{fig:intro}(a) thermal criticality is expected to be of the
Ising universality class. In Fig.~\ref{fig:ising} we present a study of the histograms of the
order parameter. We see that just above the critical $T$, $P(\psi)$
shows one peak at zero. As $T$ is lowered, the zero-peak splits into two symmetric
peaks corresponding to spontaneous symmetry breaking, just as one expects for the Ising
model. There is
no evidence for a peak at zero co-existing with the non-zero peaks,
which one would expect at a first order transition.  A study of the scaling
behavior of the $T$-dependence of the order parameter at $g=0.5$
(right panel of Fig.~\ref{fig:ising})  shows conclusive evidence that the HN order
parameter undergoes a continuous thermal Ising phase transition, as expected for its order parameter manifold. This provides our final piece of evidence that the broken
symmetry is indeed of the Haldane nematic form illustrated in Fig.~\ref{fig:intro}(a).

\begin{figure}[t!]
 \includegraphics[trim={0 0 0 0},clip, width=3.5in]{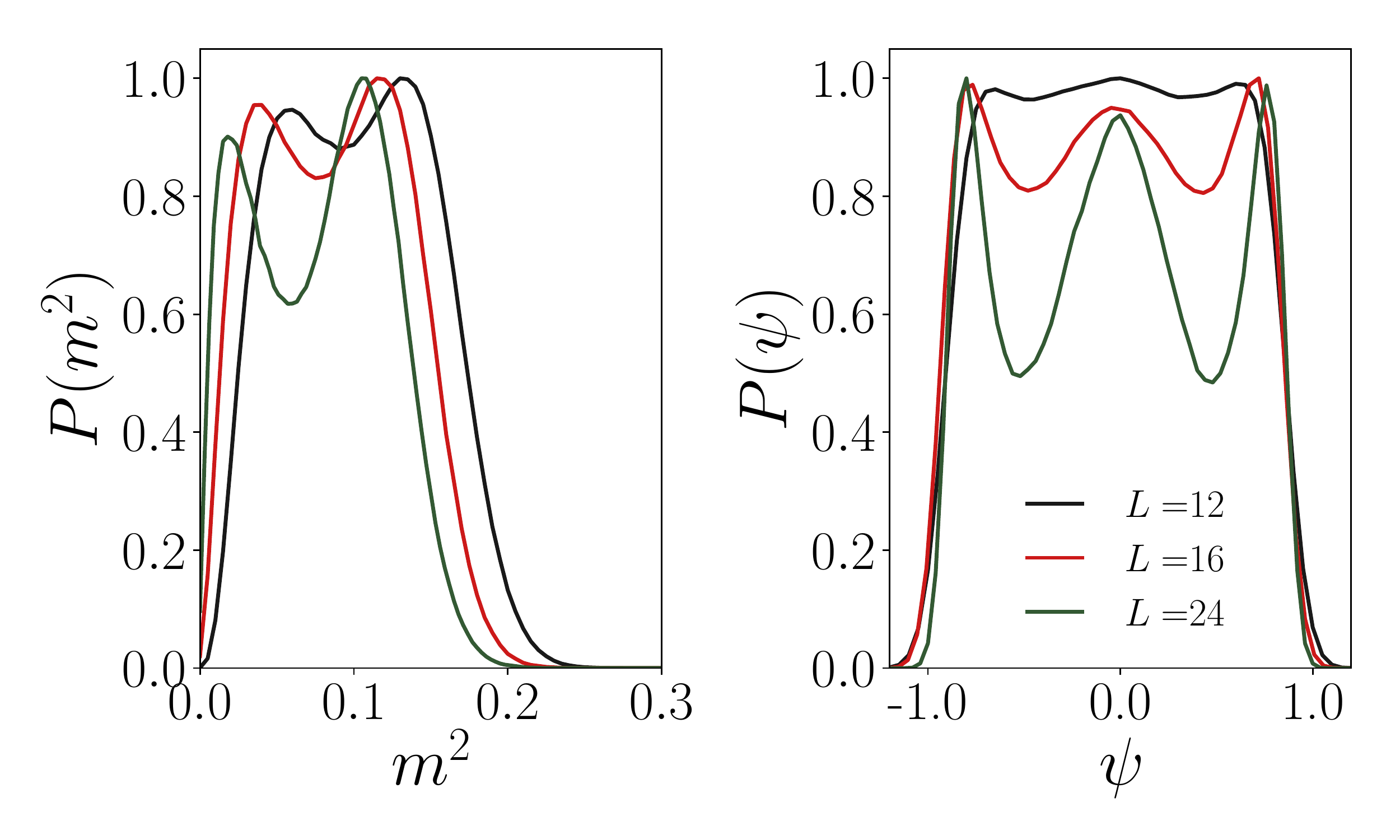} 
\caption{\label{fig:first} Evidence for first order behavior at the phase transition at $T=1$ in
  Fig.~\ref{fig:pd}. The left panel shows the
  histogram for $m^2$. The right panel shows the histograms for
  $\psi$. The data shows that the double peaked behavior clearly gets
  sharper as the system size, $L$ is increased, indicating that first
  order behavior persists in the thermodynamic limit.}
\end{figure}

A final interesting question we address is the nature of the quantum phase
transition between N\'eel-HN, labeled by a star in Fig.~\ref{fig:pd}. The field
theory for this
phase transition has been argued to be the O(4)
$\sigma$-model at topological angle $\pi$~\cite{wang2015:s1},
building on previous work for
$S=1/2$~\cite{tanaka2005:nvbs,senthil2006:topnlsm}. Very little is
known about this field theory, but a consistent scenario for a continuous transition with emergent $O(4)$ symmetry at the
critical point would require only one relevant $O(3)\times Z_2$ anisotropy that appears as the
tuning parameter $g$ in the lattice model. This delicate question has not yet been accessed in
unbiased simulations. To approach this point we
study the nature of the phase transition as we move down the thermal
phase transition line to lower temperatures. From
Fig.~\ref{fig:ising}, we have seen at
high-$T$ the transition is continuous and of the Ising type. In
Fig.~\ref{fig:first} we study data at $T=1$ (which is very low-T in the
units in which we are working) while tuning $g$ (horizontal dashed
line in Fig.~\ref{fig:pd}). The histogram data shows clear evidence
that the transition has become first order for the HN order
parameter, with a co-existence of a peak at zero (for non-HN phase) and the finite
symmetry related peaks for the HN phase. While there is no thermal
phase transition for the N\'eel order it also shows double peaks that
are incipient behavior of the first order quantum phase transition it
undergoes at $g\approx 0.17$. We thus reach the conclusion that along
the phase boundary line (solid curve in Fig.~\ref{fig:pd}) the
phase transition changes from being Ising and continuous at high-$T$ to
becoming first order at low-$T$ and remains first order at the quantum phase
transition, marked with a star. The change from continuous Ising to first order is expected to happen at a
multi-critical point somewhere along the solid line in
Fig.~\ref{fig:pd} between the two
limiting cases we have studied and is expected to be described by the
tricritical Ising field theory~\cite{cardy1996:ren}. We have not made an effort to locate this
point precisely in our phase diagram in this work.

Our finding of a first order quantum transition can be interpreted in two
different ways for the $O(4)$ sigma-model at
$\theta=\pi$. The first is simply that the field theory itself does
not have a non-trivial
critical fixed point, the other is that such a fixed point exists but it has more than one relevant
$O(3)\times Z_2$ anisotropy and thus requires more than one tuning parameter to
be reached. We note that our finding is consistent with previous
studies of the $S=1/2$ N\'eel-VBS deconfined critical point on a rectangular
lattice which is expected to be described by the same field theory and
anisotropies as
the $S=1$ N\'eel-HN studied here~\cite{haldane1988:berry,read1990:vbs,senthil2004:science} and was also found to be first order~\cite{block2013:fate}.

{\em Conclusions:} We have introduced a scheme to design general
multi-spin interactions for
spin-$S$ models without the sign problem. Our scheme opens up the
possibility to simulate a wide range of models and address the role of
$S$ on quantum phase transitions in two and
higher dimensions. Higher spin can introduce new phases not
present for $S=1/2$, including multi-polar ordered phases and new
paramagnetic phases, like the unconventional valence bond ordering we
found here and
quantum spin liquids. The theory of phase transitions between these
new phases is largely unexplored. All of these are exciting avenues for future work. 

{\em Acknowledgments:} We gratefully acknowledge useful discussion with
S.~Pujari and partial support from
NSF DMR-1611161 and Keith
B. MacAdam Graduate Excellence Fellowship. The numerical results were produced on SDSC comet
cluster through the NSF supported XSEDE award TG-DMR140061 as well as
the DLX cluster at UK.\\

\putbib[spS.bib]
\end{bibunit}

\clearpage
\begin{bibunit}[apsrev]

\onecolumngrid
\section{Supplemental Materials}
\twocolumngrid
\subsection{Split Spin Representation:}

We use the split spin representation \cite{todo,kawashima-gubernatis} to map a model of interacting spin-$S$'s onto a model of spin-$\frac{1}{2}$'s. 
In this representation spin-$S$ operators on each site are written as a sum of 2$S$ spin-$\frac{1}{2}$ operators (mini-spins) as shown in below:

\begin{equation}
 \vec{S}_i=\sum_{\mu=1}^{2S} \vec{s}^{\,a}_{i}
 \label{split-spin}
\end{equation}

The partition function of the original spin-$S$ model in terms of the resulting spin-$\frac{1}{2}$ Hamiltonian, $\tilde{H}$ can be written as: 
\begin{equation}
 Z=Tr_s(e^{-\beta \tilde{H}}\mathcal{P})
  \label{Zmini}
\end{equation}
\begin{equation}
 \mathcal{P}=\prod_i \mathcal{P}_i
 \label{projection-op}
\end{equation}

where $\mathcal{P}_i$ at each site $i$ acts on the $2^{2S}$ dimensional Hilbert space spanned by the spin-$\frac{1}{2}$'s and projects out unphysical states 
that don't belong to the spin-$S$ subspace. $\tilde{H}$ is invariant under exchange of the mini-spin indices at each site 
and therefore commutes with $\mathcal{P}$.

\subsubsection*{Projection Operator}
The projection operator has all positive matrix elements and hence can be simulated without a sign problem. Taking the spin-$1$ case for simplicity, the projection operator
is given by: 
\begin{multline}
 \mathcal{P}_i= | \uparrow \uparrow \rangle \langle \uparrow \uparrow | + |\downarrow\downarrow\rangle\langle\downarrow\downarrow| \\
 + \Big(\frac{| \uparrow \downarrow \rangle + |  \downarrow\uparrow \rangle }{\sqrt2}\Big)\Big(\frac{ \langle \uparrow \downarrow | + \langle \downarrow\uparrow|  }{\sqrt2}\Big)   
\end{multline}

\begin{figure}
  \begin{minipage}{0.22\textwidth}
    \subfigure[Moves with probability $\frac{1}{2}$]{\includegraphics[scale=0.25]{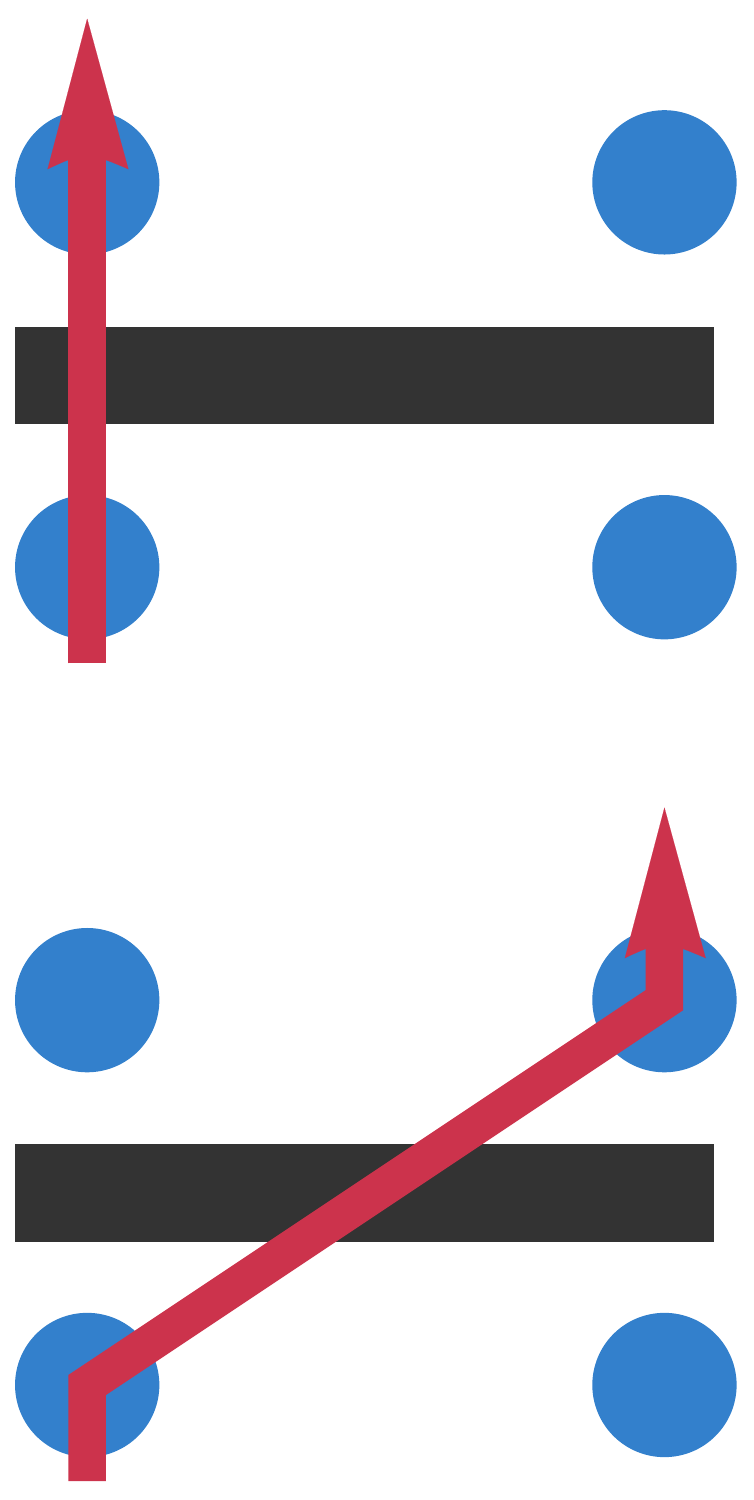}}
  \end{minipage}
  \begin{minipage}{0.22\textwidth}
    \subfigure[Moves with probability 1]{\includegraphics[scale=0.25]{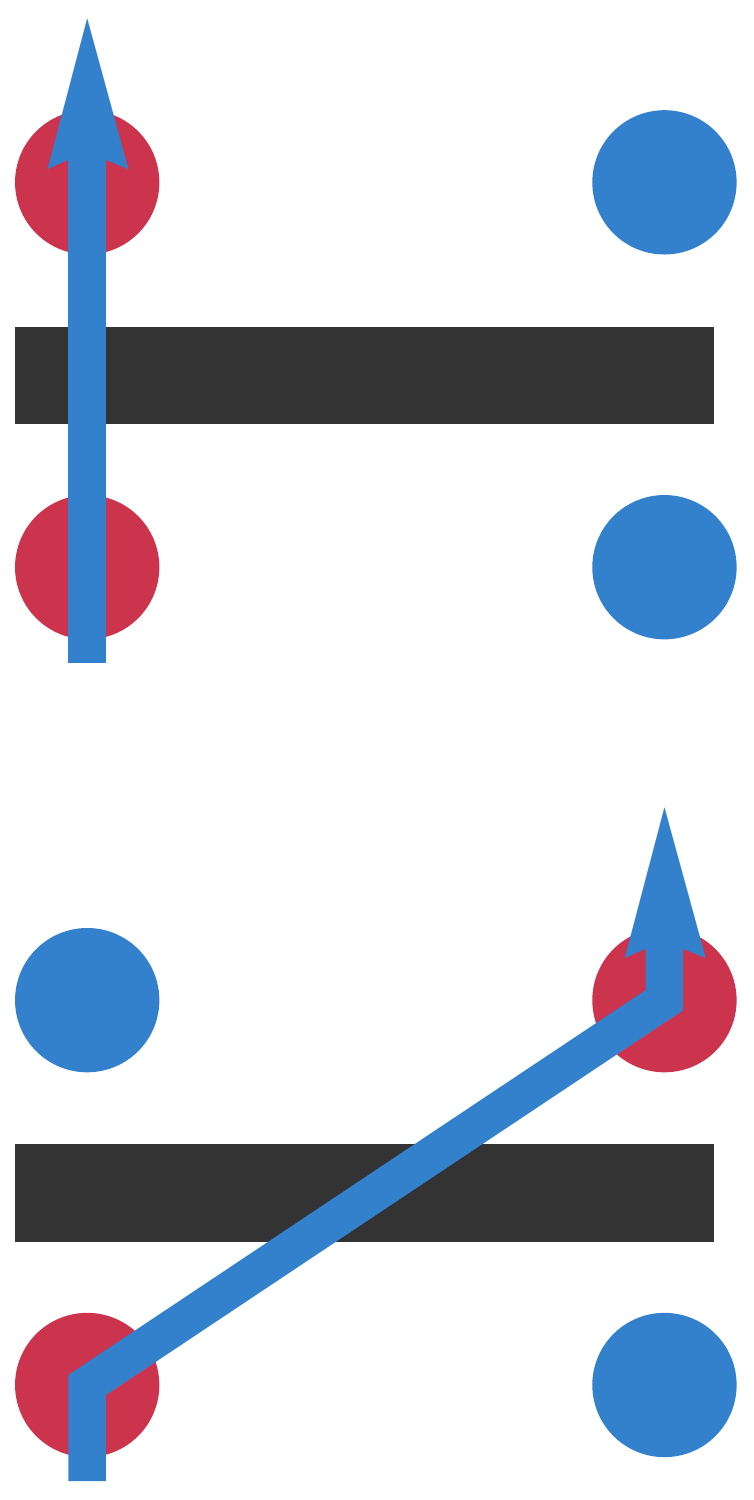}}
  \end{minipage}
  \caption{\label{proj}Loop moves to update the projection operator: The two colors represent the two spin states, a loop update flips the spin to a different
  state. (b) are equilvalent to the reverse moves of (a)}
\end{figure}

The update of this operator in our QMC procedure can be carried out using the directed loop algorithm \cite{directed-loops}. Fig. \ref{proj} shows the loop updates with their respective probabilities for this operator. 
   
The loop update in Fig. \ref{proj} can be generalized to higher spins as follows: the loop entering the operator continues in the same direction exiting on 
any mini-spin who's state is the same as that of the mini-spin at which it entered. The operator can also be thought of as a ``soft'' boundary condition in the 
imaginary time direction. The spin-$S$ state is the highest spin that can be gotten from the sum on 2S spin-$\frac{1}{2}$'s and the highest spin state 
wavefunction has to be completely symmetric in the mini-spins. Hence, the $\mathcal{P}_i$ operator is identical to a local symmetrization 
operator at site $i$. It ensures that the Hamiltonian operators propagate the spin state in the imaginary time $(\tau)$ direction so that the spin state 
at $\tau=\beta$ is the spin state at $\tau=0$ upto a permutation of the $2S$ mini-spins at each site.

\subsection{Designer Hamiltonians:}

Consider the spin-$S$ nearest neighbour Heisenberg Antiferromagnet (upto a constant, $-S^2$) on a bipartite lattice,
\begin{equation}
  \begin{split}
   H &=-J \sum_{\langle ij \rangle} (S^2-\vec{S}_i. \vec{S}_j) \\
     &= J\sum_{\langle ij \rangle} H_{ij}
  \end{split}
 \label{spinSham}
\end{equation}

where 

\begin{equation}
 H_{ij}=\vec{S}_i. \vec{S}_j-S^2
\end{equation}

By carrying out a Unitary transformation of the spin operators on one of the sublattices such that $S^+_i \rightarrow -S^-_i$
and $S^-_i \rightarrow -S^+_i$, one can easily see that the operator $H_{ij}$ has all negative matrix
elements. This can be simulated without a sign problem using QMC. This is because the Hamiltonian enters the series expansion of the partition function as
powers of $-\beta H$ giving a positive probability for each term in the partition function \cite{sandvik}. For the spin-$\frac{1}{2}$ case this quantity $P_{ij}=\frac{1}{4}-\vec{S}_i. \vec{S}_j$ is a singlet projection operator. All models constructed out of $P_{ij}$, e.g. products of 
$P_{ij}$ on two bonds on a plaquette, are sign problem-free.
We use this fact to construct designer Hamiltonians for arbitrary $S$ as described in the main manuscript. \\

In terms of spin-$\frac{1}{2}$'s $H_{ij}$ can be written as:
\begin{equation}
 \tilde{H}_{ij} =-J \sum_{a,b=1}^{2S} \Big(\frac{\mathbbm{1}}{4}-\vec{s}_i^{\, a}. \vec{s}_j^{\, b}\Big)
 \label{heisspinhalf}
\end{equation}

\begin{figure}
 \begin{minipage}{0.22\textwidth}
  \subfigure{\includegraphics[scale=0.16]{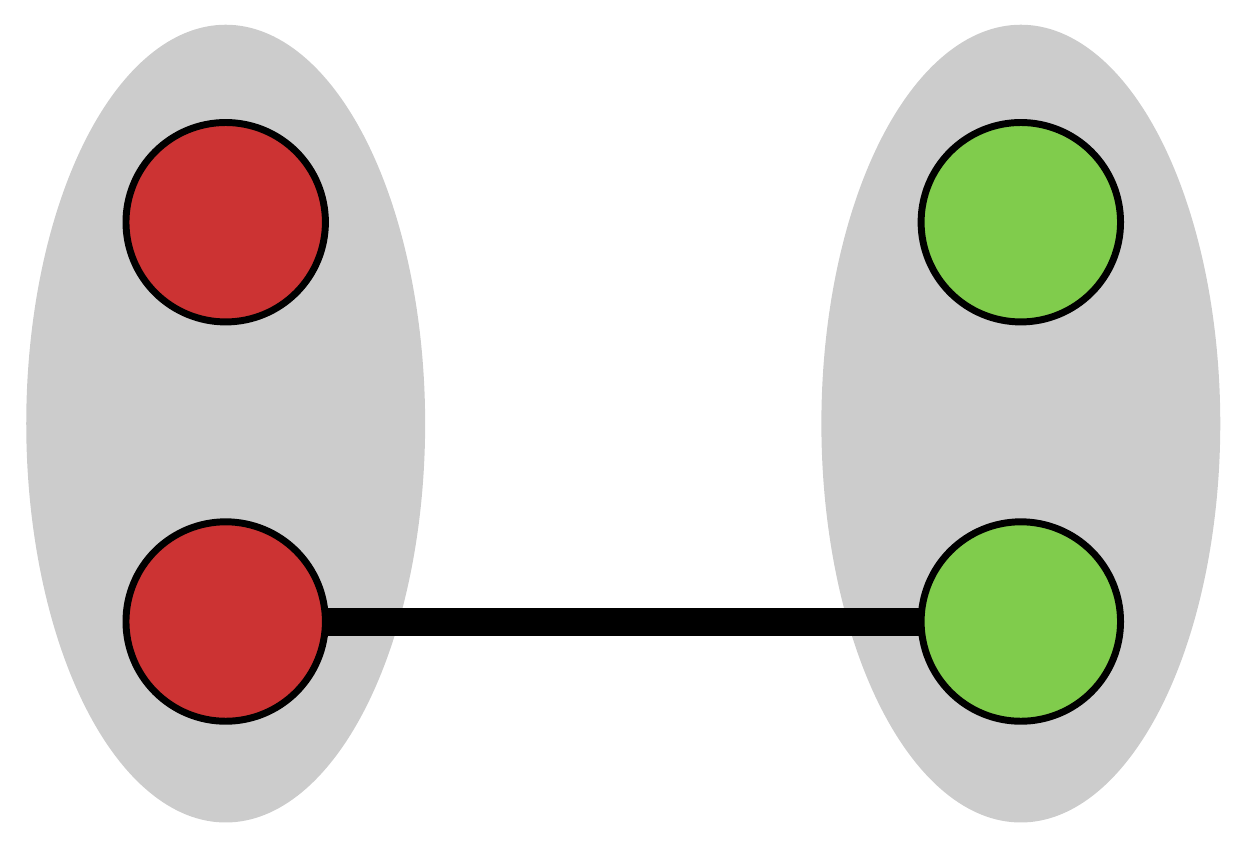}}
  \vspace*{2mm}
  \subfigure{\includegraphics[scale=0.16]{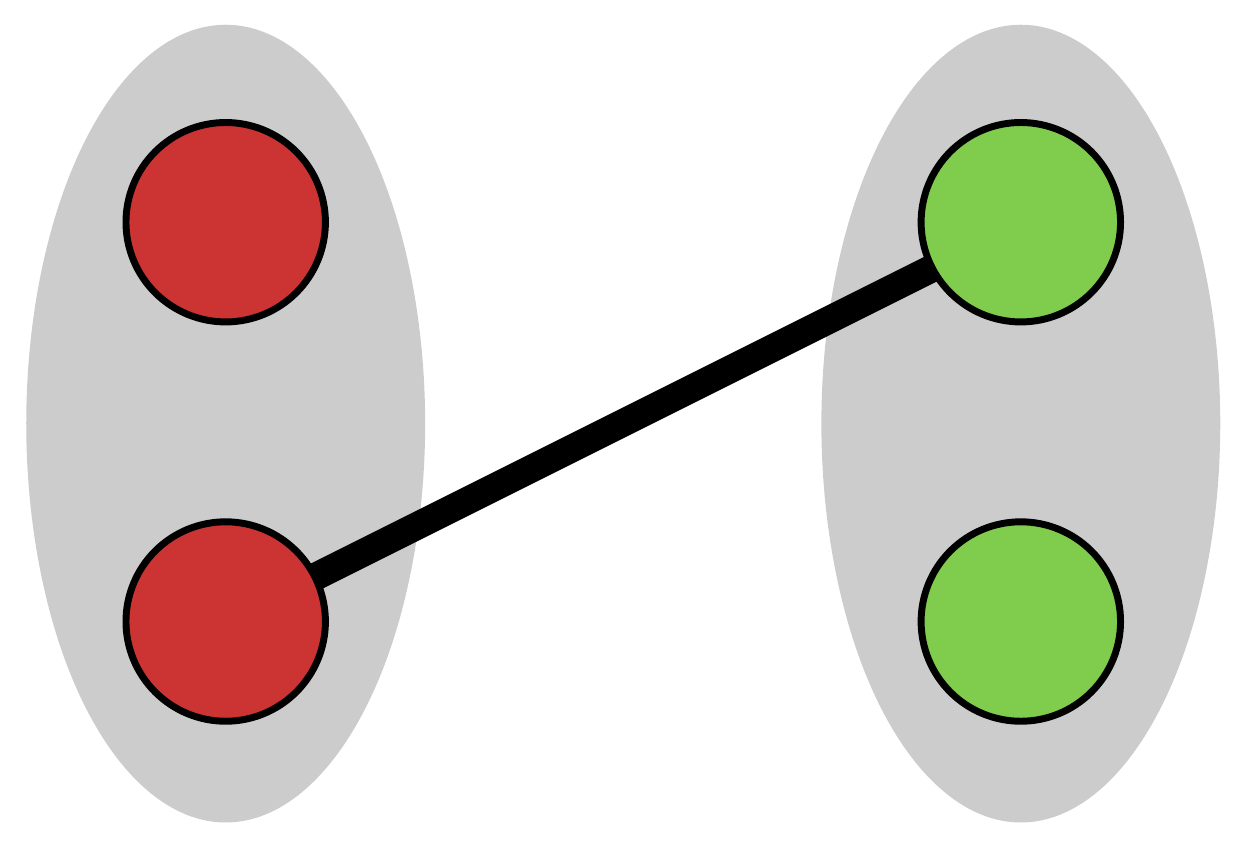}} 
 \end{minipage}
 \begin{minipage}{0.22\textwidth}
  \subfigure{\includegraphics[scale=0.16]{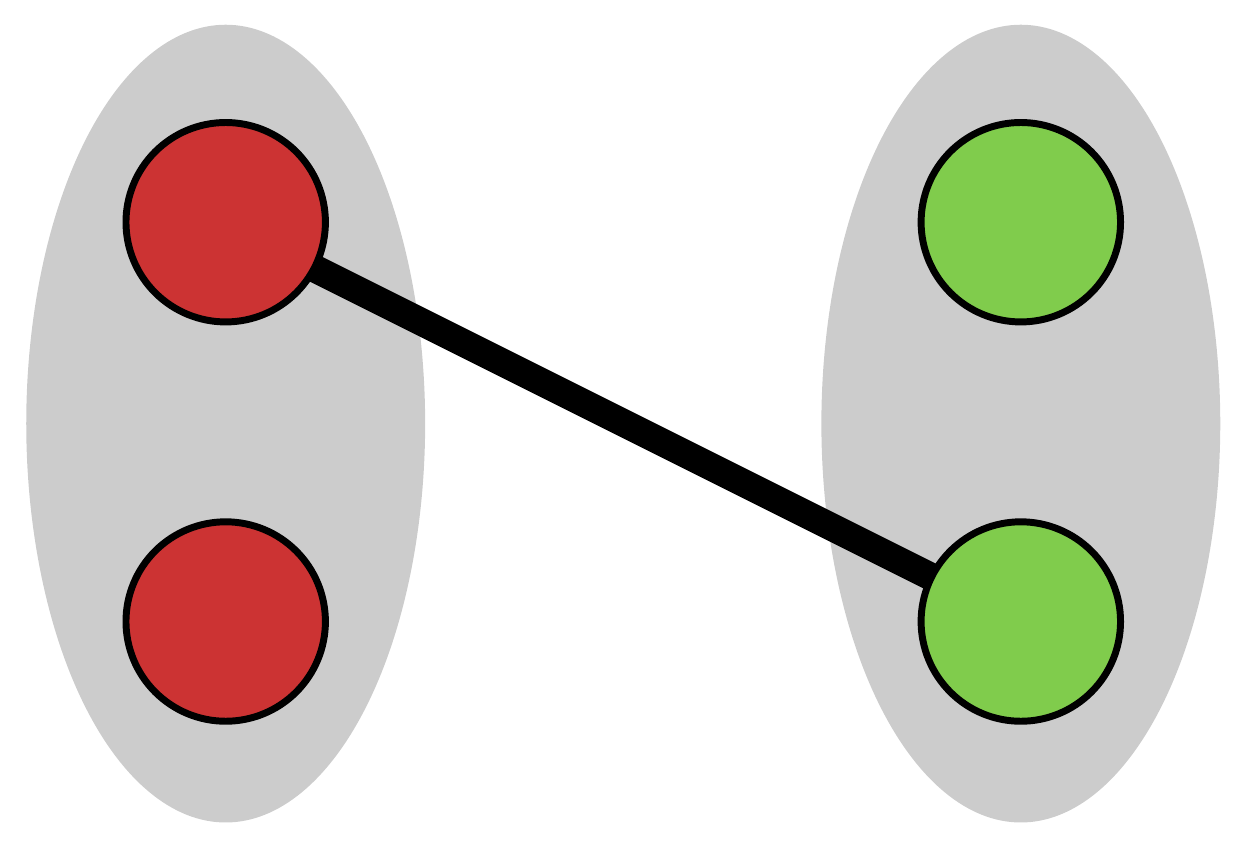}}
  \vspace*{2mm}
  \subfigure{\includegraphics[scale=0.16]{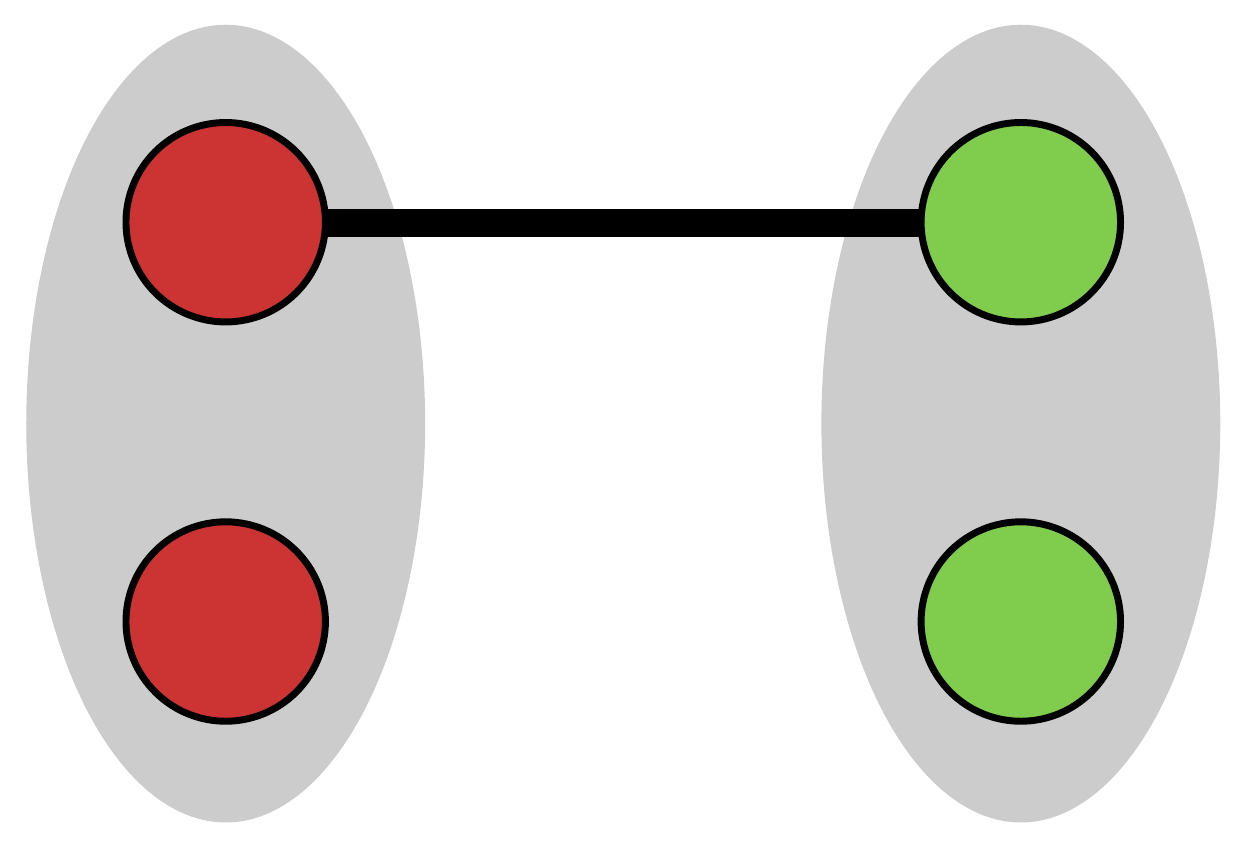}} 
 \end{minipage}
 \caption{\label{heis} Each gray bubble with two circles represents two spin-$\frac{1}{2}$'s at each spin-1 site. The black line is the operator $P_{ij}$ acting on the two mini-spins. The four diagrams denote the four terms in the summation in Eq. \ref{heisspinhalf}}
\end{figure}

\begin{figure}
 \subfigure{\includegraphics[scale=0.16]{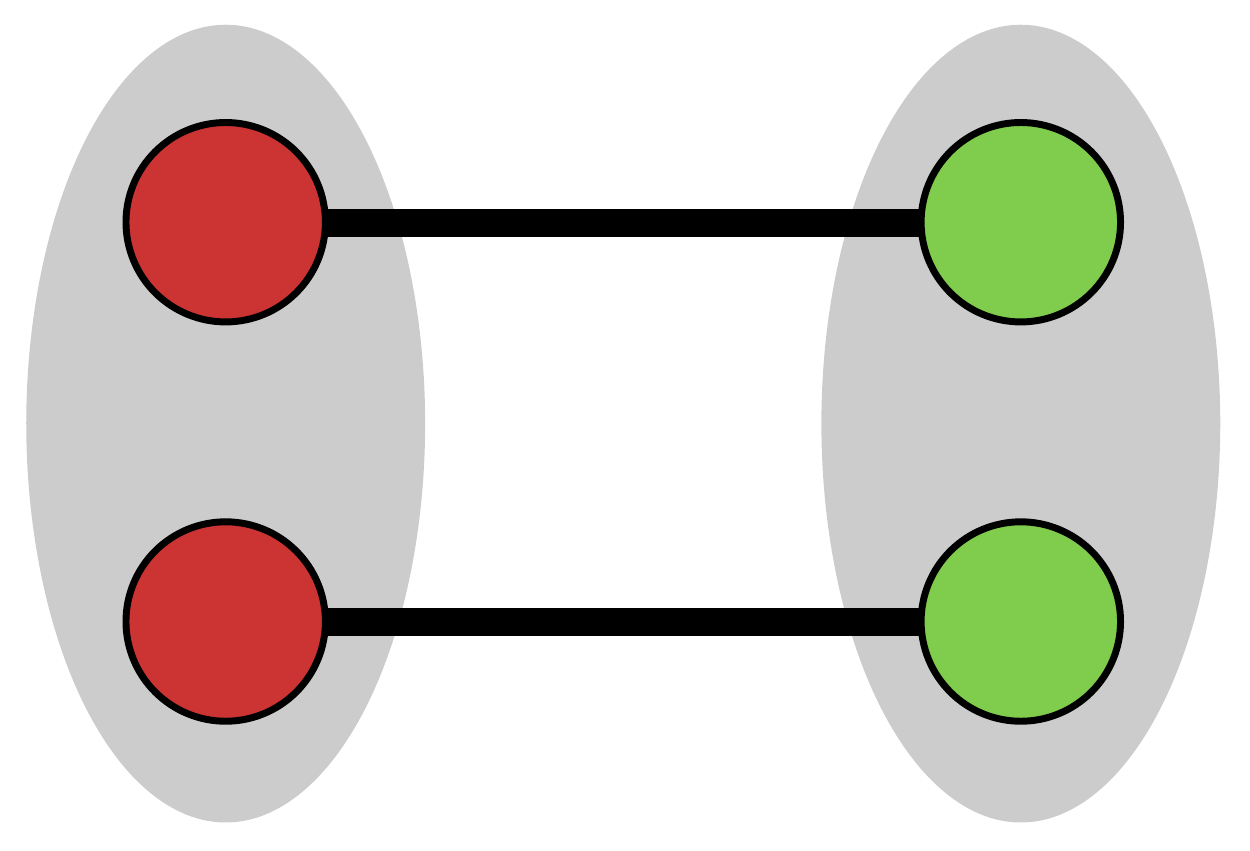}}
 \hspace*{5mm}
 \subfigure{\includegraphics[scale=0.16]{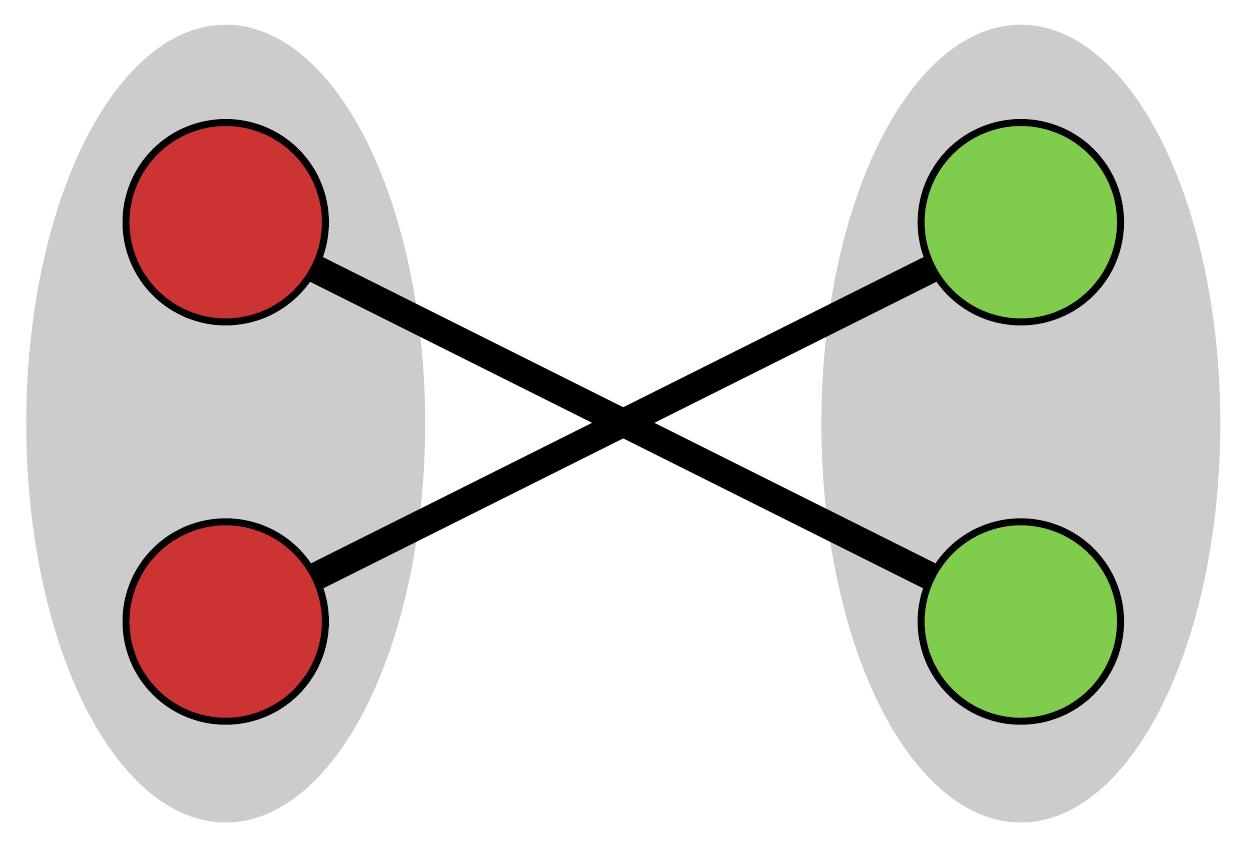}}
 \caption{\label{biquadminifig} These two diagrams is equivalent to the first two terms in the square bracket in Eq. \ref{biquadmini}}
\end{figure}

The above spin-$\frac{1}{2}$ Hamiltonian is a sum of four terms, each of which can be depicted pictorially as in Fig. \ref{heis}. Similarly, the biquadratic term given by $H_{biquad}$ in Eq. \ref{biquad} can be expressed in terms of the mini-spins as $\tilde{H}_{biquad}$  as in 
equation \ref{biquadmini}. The last two terms colored in red in Eq. \ref{biquadmini} get killed by the action of the projection operator $\mathcal{P}$ at each of these sites. This is because these operators are singlet projection operators on the two mini-spins at the same site.
This operator is anti-symmetric in these two mini-spins and therefore gets cancelled by the symmetrization operator $\mathcal{P}$. The remaining two terms in Eq. \ref{biquadmini} can be understood more easily with the help of Fig. \ref{biquadminifig}

\begin{equation}
 H_{biquad}=(\vec{S}_i.\vec{S}_j)^2-\mathbbm{1}
 \label{biquad}
\end{equation}

\begin{multline}
 \tilde{H}_{biquad}=\sum_{a,b=1}^2 \Big[\Big(\frac{\mathbbm{1}}{4}-\vec{s}^{\, a}_i.\vec{s}^{\, a}_j\Big).\Big(\frac{\mathbbm{1}}{4}-\vec{s}^{\, b}_i.\vec{s}^{\, b}_j\Big)+\\ \Big(\frac{\mathbbm{\mathbbm{1}}}{4}-\vec{s}^{\, a}_i.\vec{s}^{\, b}_j\Big).\Big(\frac{\mathbbm{1}}{4}-\vec{s}^{\, b}_i.\vec{s}^{\, a}_j\Big)+ \\ \textcolor{red} {\frac{1}{2} \Big(\vec{s}^{\, a}_i.\vec{s}^{\, b}_i - \frac{\mathbbm{1}}{4} \Big) + \frac{1}{2} \Big(\vec{s}^{\, a}_j.\vec{s}^{\, b}_j - \frac{\mathbbm{1}}{4} \Big)} \Big]
 \label{biquadmini}
\end{multline}

\begin{figure}
  \begin{minipage}{0.22\textwidth}
    \subfigure{\includegraphics[scale=0.26]{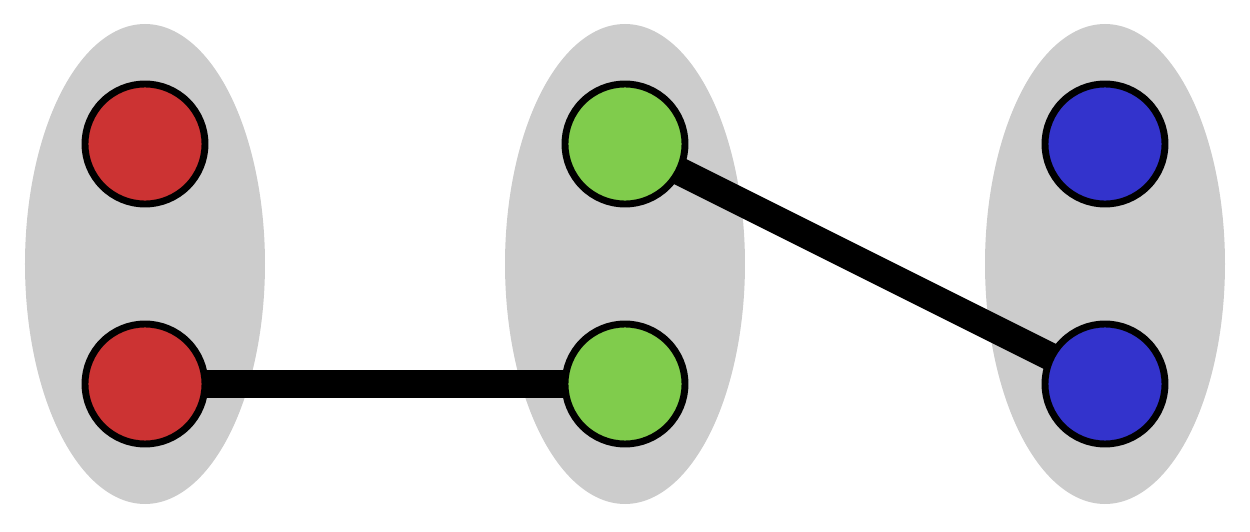}}
    \vspace*{2mm}
    \subfigure{\includegraphics[scale=0.26]{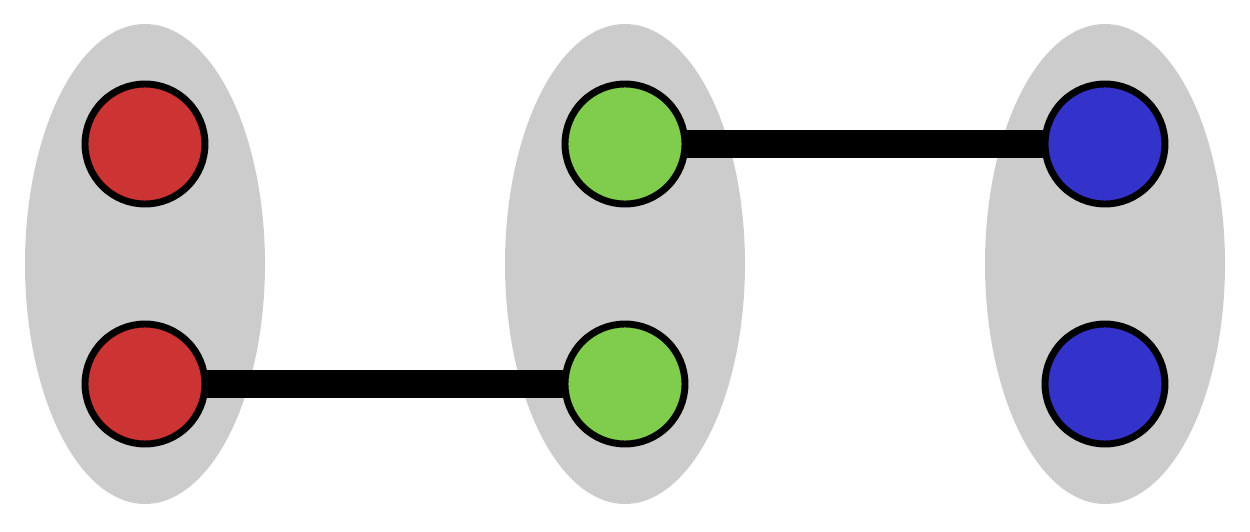}}
    \vspace*{2mm}
    \subfigure{\includegraphics[scale=0.26]{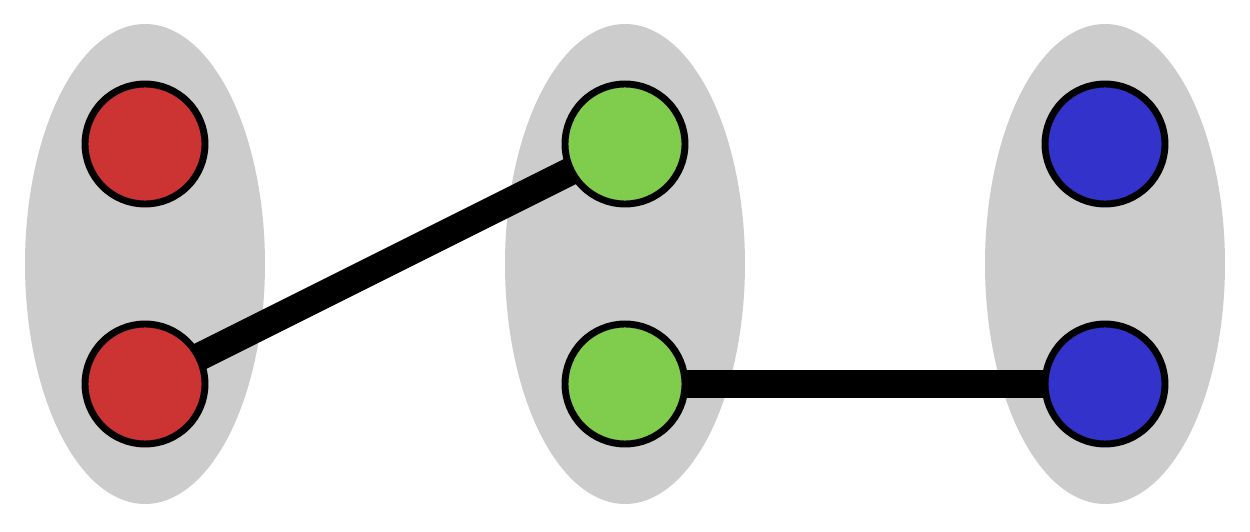}}
    \vspace*{2mm}
    \subfigure{\includegraphics[scale=0.26]{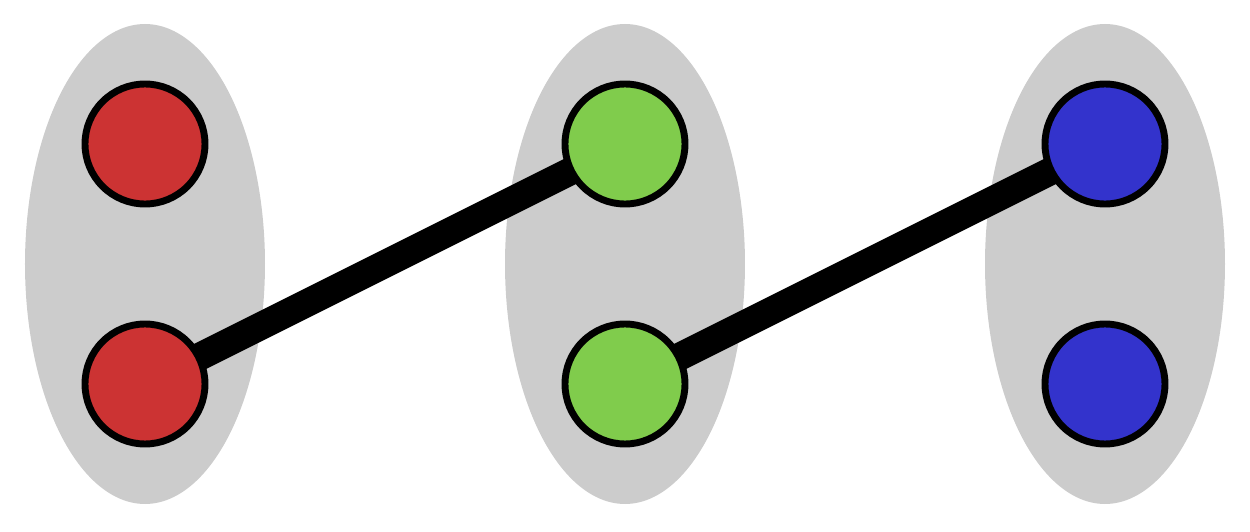}}
  \end{minipage}
  \begin{minipage}{0.22\textwidth}
    \subfigure{\includegraphics[scale=0.26]{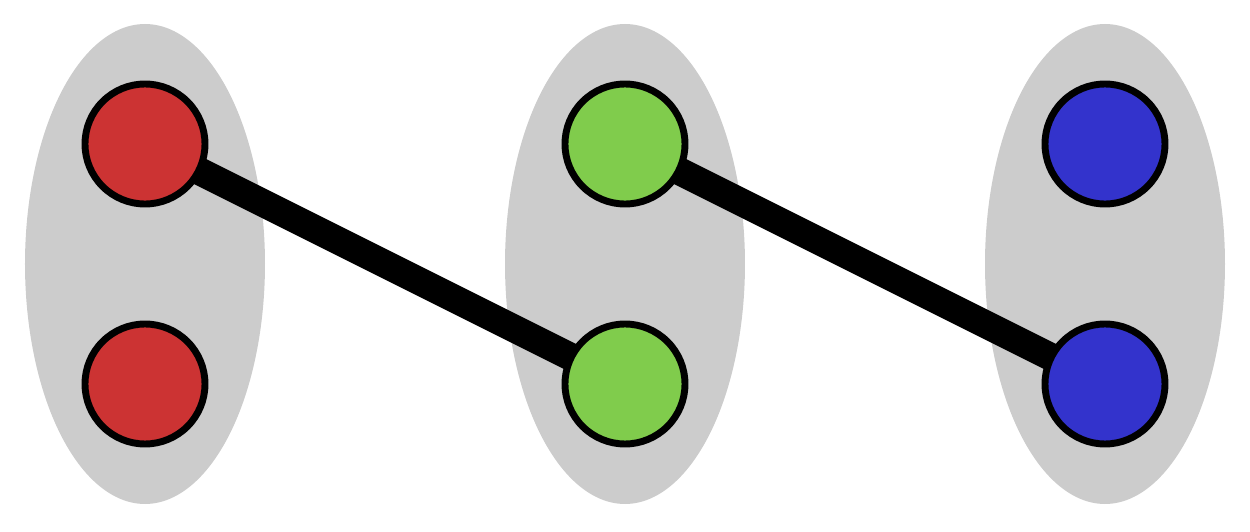}}
    \vspace*{2mm}
    \subfigure{\includegraphics[scale=0.26]{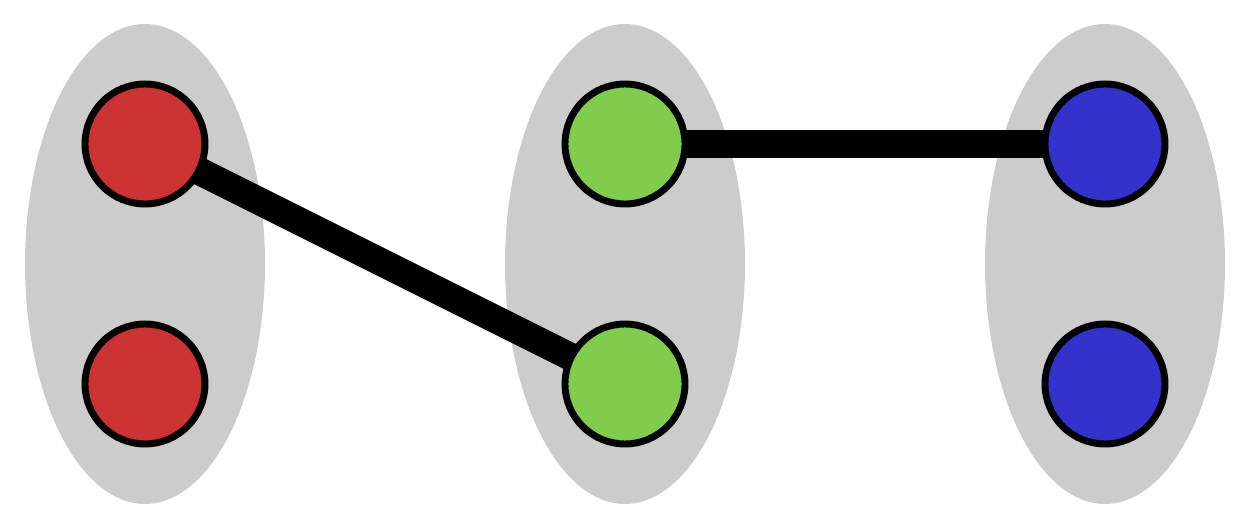}}
    \vspace*{2mm}
    \subfigure{\includegraphics[scale=0.26]{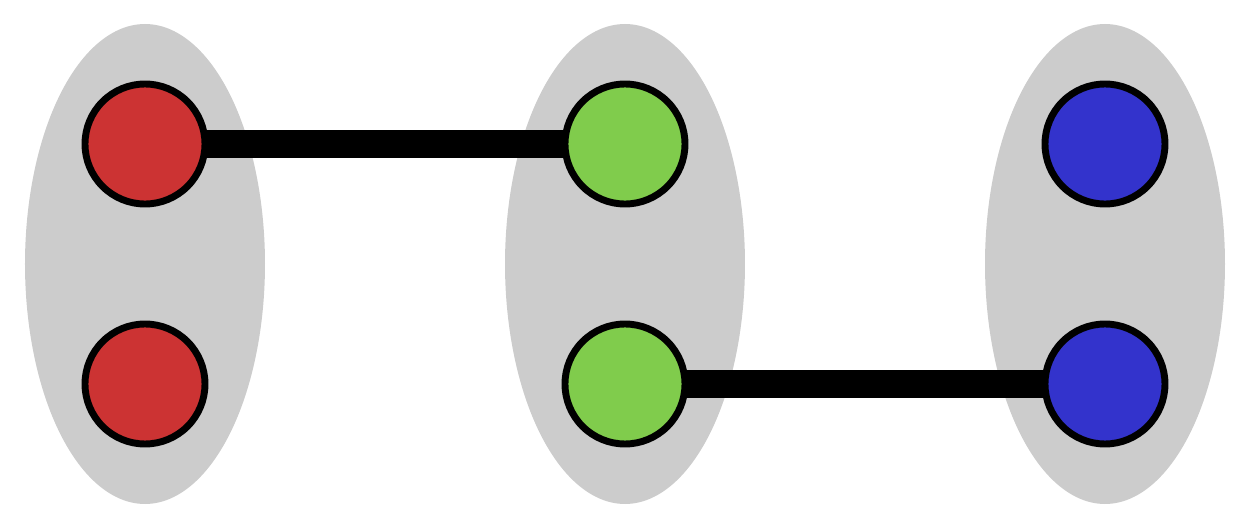}}
    \vspace*{2mm}
    \subfigure{\includegraphics[scale=0.26]{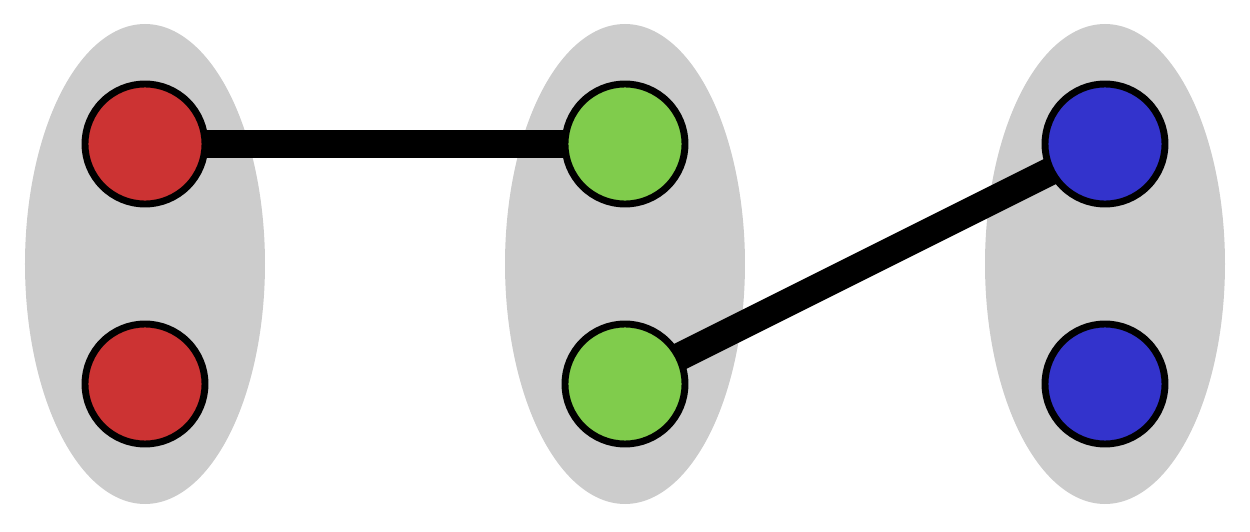}}
  \end{minipage}
  \caption{\label{splitspin1}Interaction between three spin-1's in terms of the mini-spins}
\end{figure}

We now construct a sign problem-free interaction symmetric in the mini-spins at each site, directly in the spin-$\frac{1}{2}$ language. This simple construction involves three neighbouring sites. A singlet projection operator acts on each pair of neighbouring spin-$\frac{1}{2}$'s on the three sites, such that none of these singlets touch. 
The 8 possible ways to form singlets on three sites in this manner are shown in Fig. \ref{splitspin1}

\vspace*{5mm}

The Hamiltonian described by Fig. \ref{splitspin1} is written down in Eq. \ref{Htildeint}. 

\begin{equation}
 \tilde{H}^{ijk}_{3} = -\sum_{a,d,b \neq c} \Big(\frac{\mathbbm{1}}{4}-\vec{s}^{\, a}_i.\vec{s}^{\, b}_j\Big)\Big(\frac{\mathbbm{1}}{4}-\vec{s}^{\, c}_j.\vec{s}^{\, d}_k\Big) + h.c.
 \label{Htildeint}
\end{equation}

\begin{figure}
  \begin{minipage}{0.22\textwidth}
    \subfigure{\includegraphics[scale=0.26]{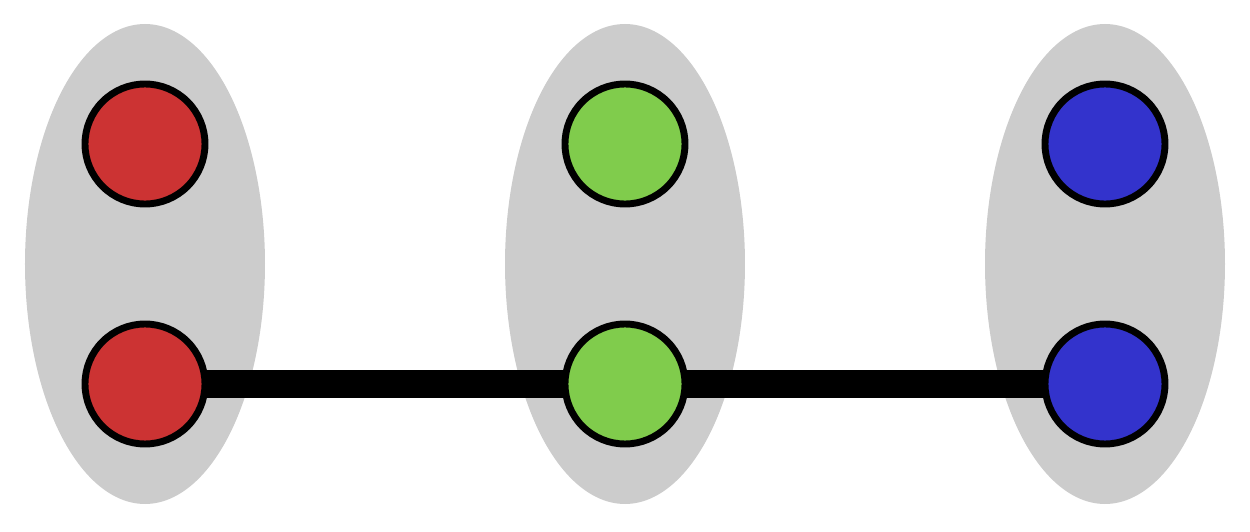}}
    \vspace*{2mm}
    \subfigure{\includegraphics[scale=0.26]{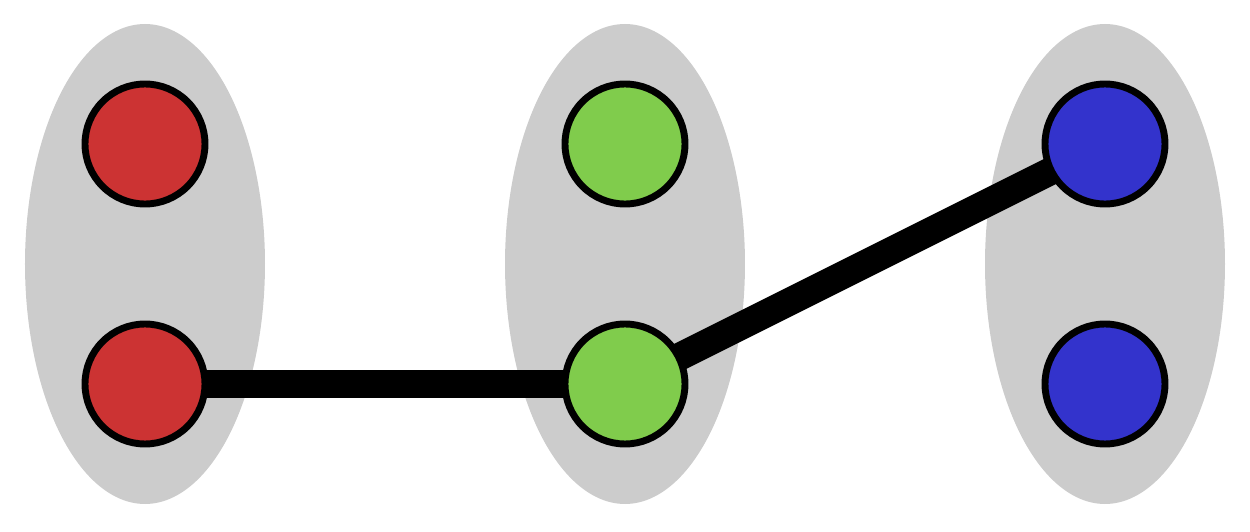}}
    \vspace*{2mm}
    \subfigure{\includegraphics[scale=0.26]{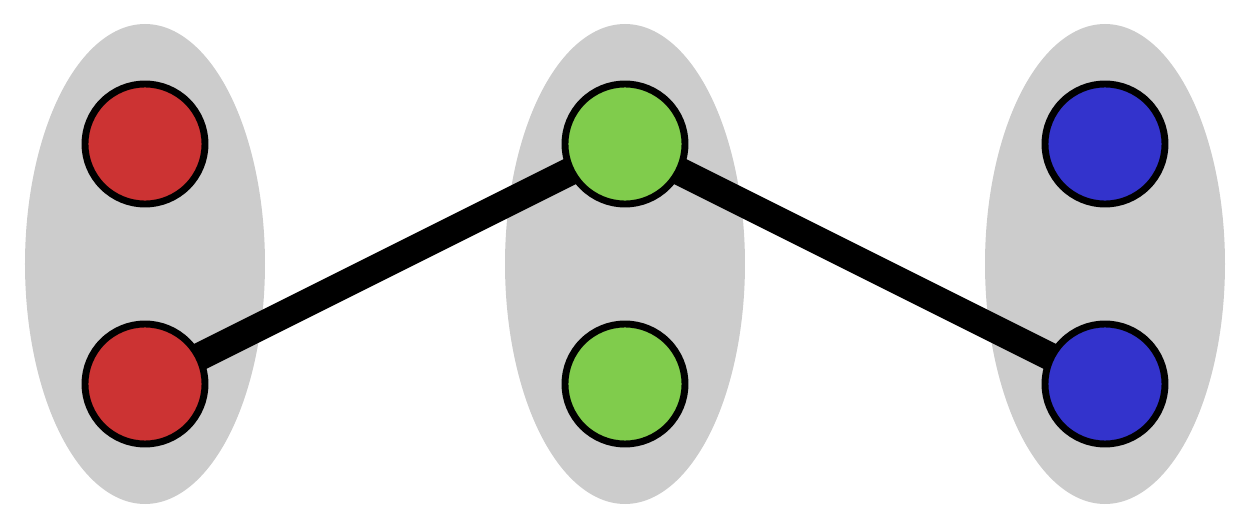}}
    \vspace*{2mm}
    \subfigure{\includegraphics[scale=0.26]{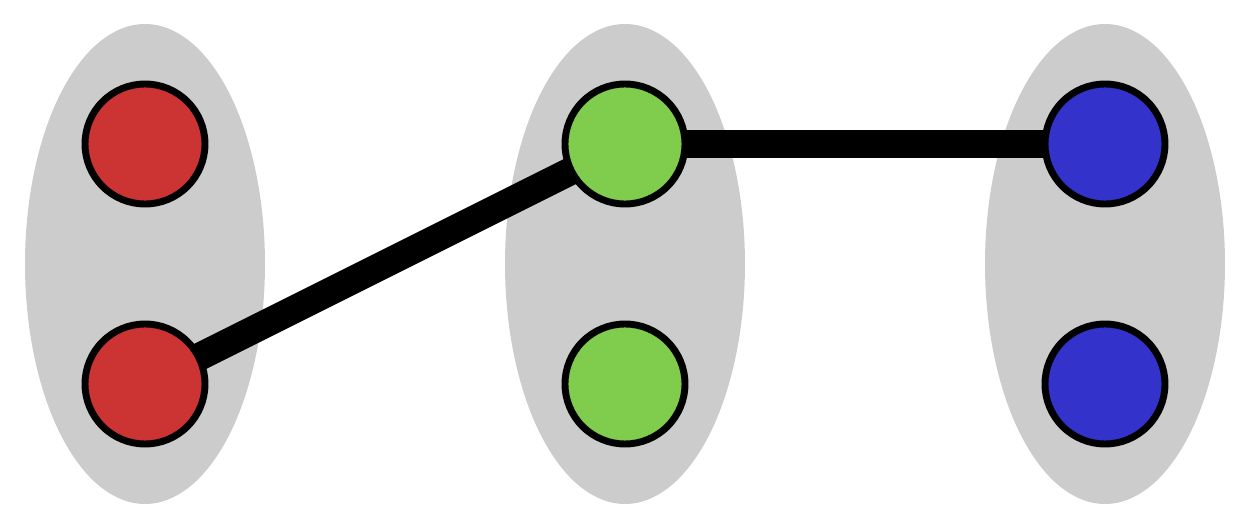}}
  \end{minipage}
  \begin{minipage}{0.22\textwidth}
    \subfigure{\includegraphics[scale=0.26]{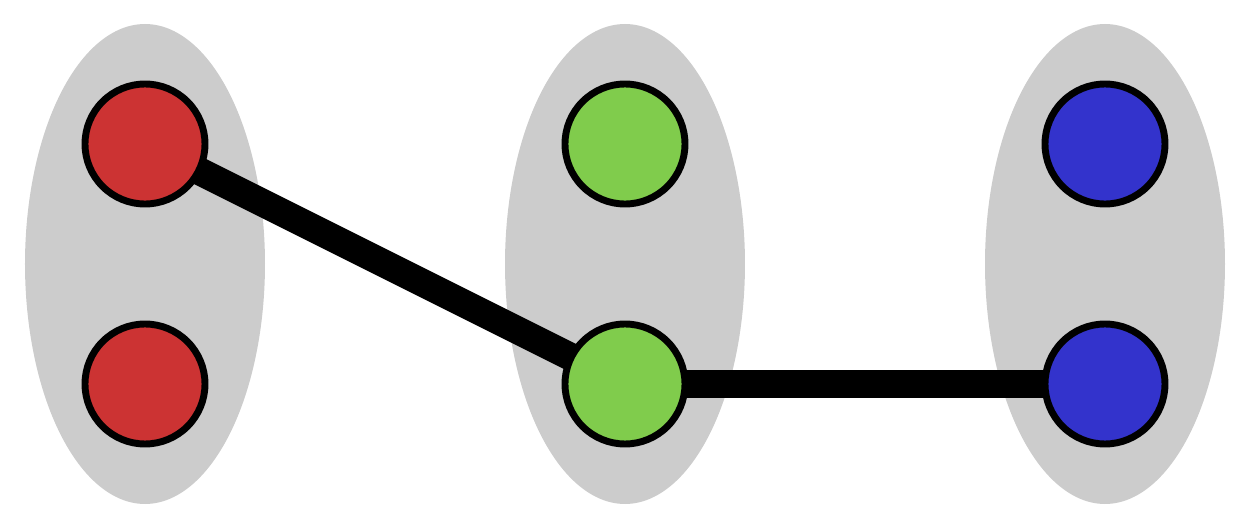}}
    \vspace*{2mm}
    \subfigure{\includegraphics[scale=0.26]{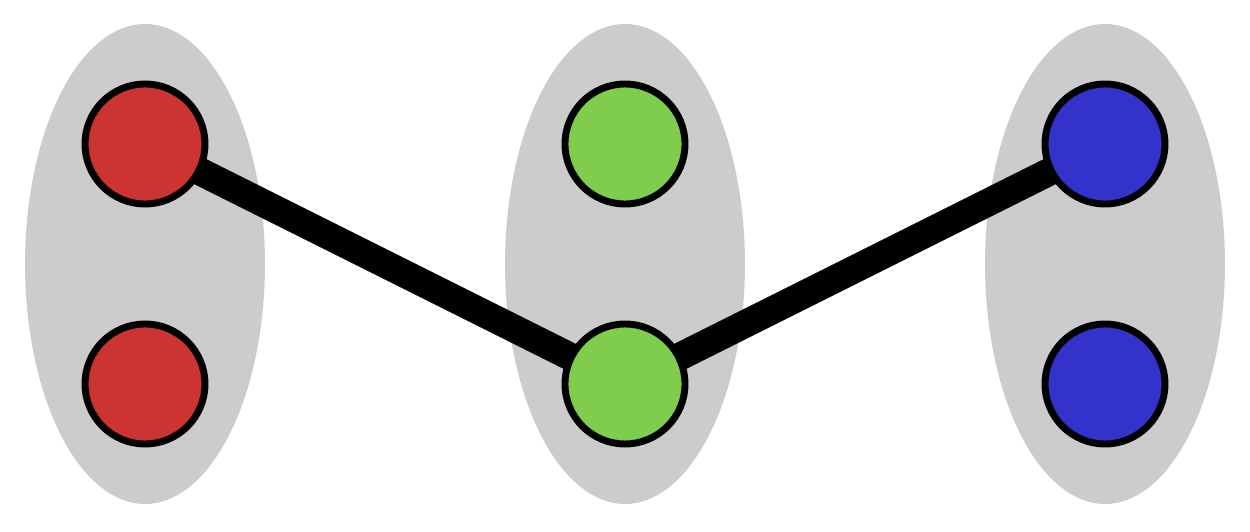}}
    \vspace*{2mm}
    \subfigure{\includegraphics[scale=0.26]{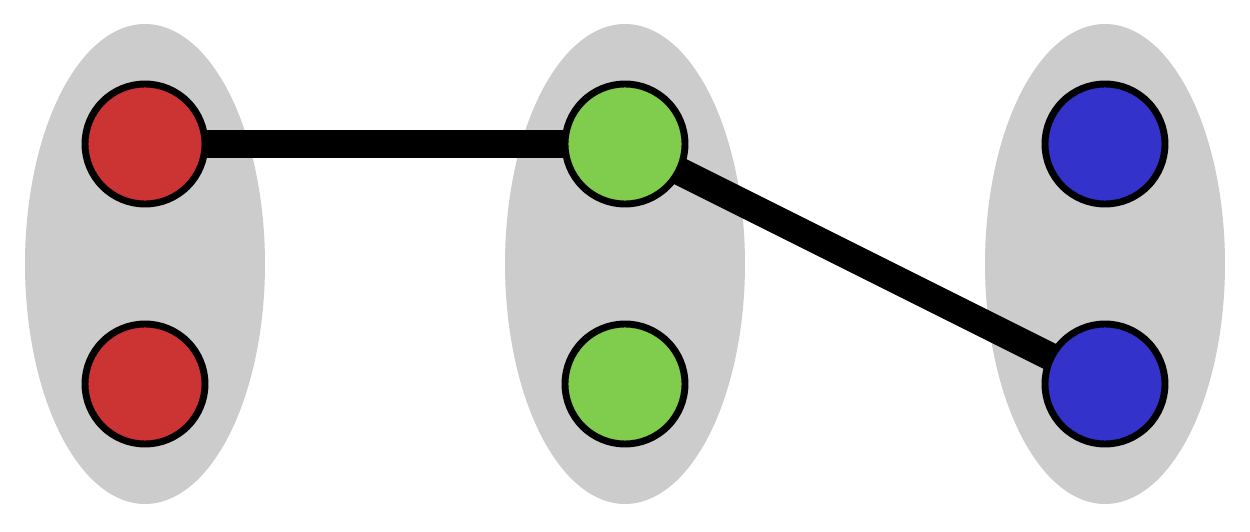}}
    \vspace*{2mm}
    \subfigure{\includegraphics[scale=0.26]{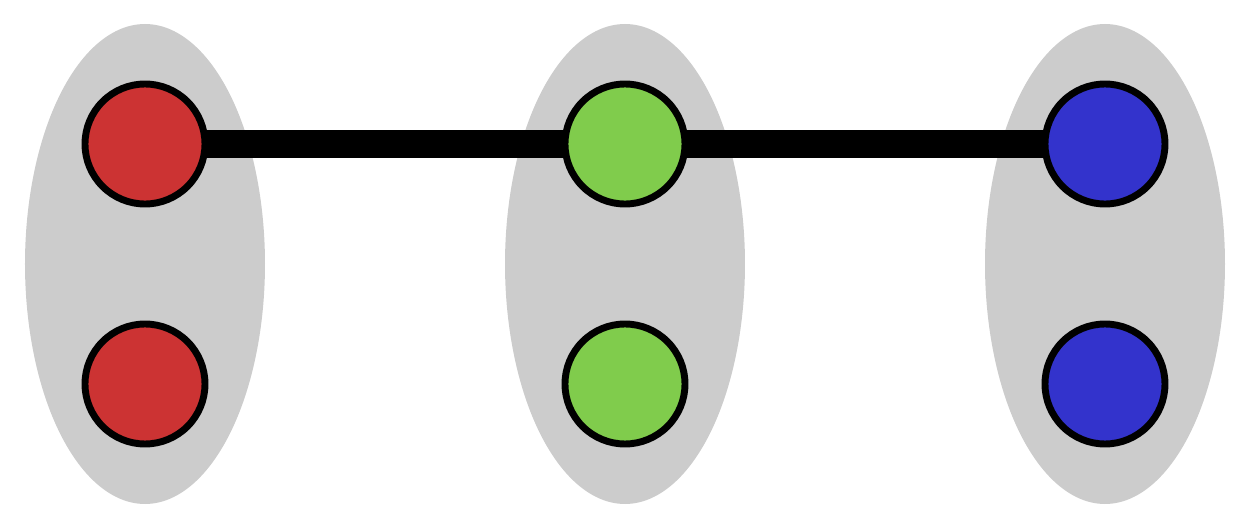}}
  \end{minipage}
  \caption{\label{splitspin2}Interactions involving three spin-$\frac{1}{2}$'s: the middle spin squares to 1 resulting in a net two spin interaction}
\end{figure}

This interaction always involves four spin-$\frac{1}{2}$'s. Interactions involving three spin-$\frac{1}{2}$'s like the ones shown in Fig. \ref{splitspin2}
can be shown to reduce to two spin interactions.

\vspace*{5mm}

We now proceed to work out this interaction in terms of the original spin-1 operators.
\begin{equation} 
\label{calculation}
\begin{split}
\tilde{H}^{ijk}_{3} & = -\sum_{a,d,b \neq c} \Big(\frac{\mathbbm{1}}{4}-\vec{s}^{\, a}_i.\vec{s}^{\, b}_j\Big)\Big(\frac{\mathbbm{1}}{4}-\vec{s}^{\, c}_j.\vec{s}^{\, d}_k\Big) \\
 & = -\Big[ \sum_{a,b,c,d} \Big(\frac{\mathbbm{1}}{4}-\vec{s}^{\, a}_i.\vec{s}^{\, b}_j\Big)\Big(\frac{\mathbbm{1}}{4}-\vec{s}^{\, c}_j.\vec{s}^{\, d}_k\Big) \\
 & -  \sum_{a,b,c} \Big(\frac{\mathbbm{1}}{4}-\vec{s}^{\, a}_i.\vec{s}^{\, b}_j\Big)\Big(\frac{\mathbbm{1}}{4}-\vec{s}^{\, b}_j.\vec{s}^{\, c}_k\Big)\Big]
\end{split}
\end{equation}

In the square bracket in the last line of Eq. \ref{calculation}: 
\begin{enumerate}[label=\alph*)]
 \item the first term  is the sum of all the terms in Fig. \ref{splitspin1} and Fig. \ref{splitspin2} 
 \item the second term is a sum on all the terms in Fig. \ref{splitspin2} 
\end{enumerate}
The two terms  are not individually Hermitian, so they are first added to their corresponding Hermitian conjugates before simplifying to get 
$H^{d,ijk}_{3}$ and $H^{d',ijk}_{3}$ respectively in Eq. \ref{Hint1} .  

\begin{equation}
 \label{Hint1}
 \begin{split}
  H^{d,ijk}_{3}&= - (\mathbbm{1} - \vec{S}_i.\vec{S}_j)(\mathbbm{1} - \vec{S}_j.\vec{S}_k) + h.c.\\
 H^{d',ijk}_{3}&= (\vec{S}_i.\vec{S}_j+\vec{S}_j.\vec{S}_k-\vec{S}_i.\vec{S}_k-\mathbbm{1})
 \end{split} 
\end{equation}

Finally, our constructed three spin interaction in terms of the spin-1's 

\begin{equation}
\label{Hint2}
\begin{split}
 H^{ijk}_{3}&=H^{d,ijk}_{3}+H^{d',ijk}_{3} \\ 
        &= -\vec{S}_i.\vec{S}_j \vec{S}_j.\vec{S}_k+\frac{1}{2}(\vec{S}_i.\vec{S}_j+\vec{S}_i.\vec{S}_k+\vec{S}_j.\vec{S}_k+2)+h.c.
\end{split}
\end{equation}

as in Eq. 3, 4 in the main manuscript.

\subsection{Measurements} 
\label{sec:measurements}

Here we outline the order parameters that we used to characterize the different phases:
\begin{enumerate}
 \item The spin spin correlation function is used to identify the magnetic order. The Fourier transform of $\langle S^z(\vec{0})S^z(\vec{r})\rangle$  has a Bragg peak at the $\vec{k}=(\pi,\pi)$, the height of this peak is our order parameter $\langle m^2 \rangle$\footnote{We define our Fourier transforms so that our order parameters are intensive}.
 \item The spin stiffness defined by Eq. \ref{stiff} is another quantity used to detect the magnetic phase
 \begin{equation}
  \rho_s=\frac{\partial^2 E(\phi)}{\partial \phi^2}\bigg|_{\phi=0} \\
  \label{stiff}
 \end{equation}
Here E($\phi$) is the energy of the system when you add a twist of $\phi$ in the boundary condition in either the $x$ or the $y$ direction. In the QMC, this quantity is related to the winding number of loops in the direction that the twist has been added:
\begin{equation}
  \rho_s= \frac{\langle W^2 \rangle}{\beta}\\
  \label{winding}
 \end{equation}
 where $\beta$ is the inverse temperature

 \item The correlation function of the singlet projection operator between neighbouring spins, $B_{i}(\vec{r})$ (Eq. \ref{sing-proj}), 
 helps determine the presence of the Valence Bond Solid (VBS) order.
 \begin{equation}
  B_{i}(\vec{r})=\sum_{a,b}\Big(\frac{1}{4}-\vec{s}^{\, a}_{\vec{r}}.\vec{s}^{\, b}_{\vec{r}+i}\Big) \\
  \label{sing-proj}
 \end{equation}
 where $i=\hat{x}$ or $\hat{y}$. 
 A Bragg peak in the Fourier transform of $J \langle B_{i}(\vec{0})B_{i}(\vec{r})\rangle$ (where $J$ is the Heisenberg coupling) at $\vec{k}=(\pi,0)$ or $\vec{k}=(0,\pi)$ 
 indicates VBS order on the square lattice. The height of this peak is the VBS order parameter given by $\langle \phi^2 \rangle$ 
 \item The Haldane Nematic phase is characterized by long range order in the quantity $\psi(\vec{r})$ which is locally defined at a site as 
 $\psi(\vec{r})=B_{\hat{x}}(\vec{r})-B_{\hat{y}}(\vec{r})$. We define our order parameter for this phase by the height of the Bragg peak  in the 
 Fourier transform of $J \langle \psi(0) \psi(\vec{r})\rangle$ (where $J$ is the Heisenberg coupling) at the $\vec{k}=(0,0)$.

 \begin{figure}
  \centering
  \includegraphics[scale=0.4]{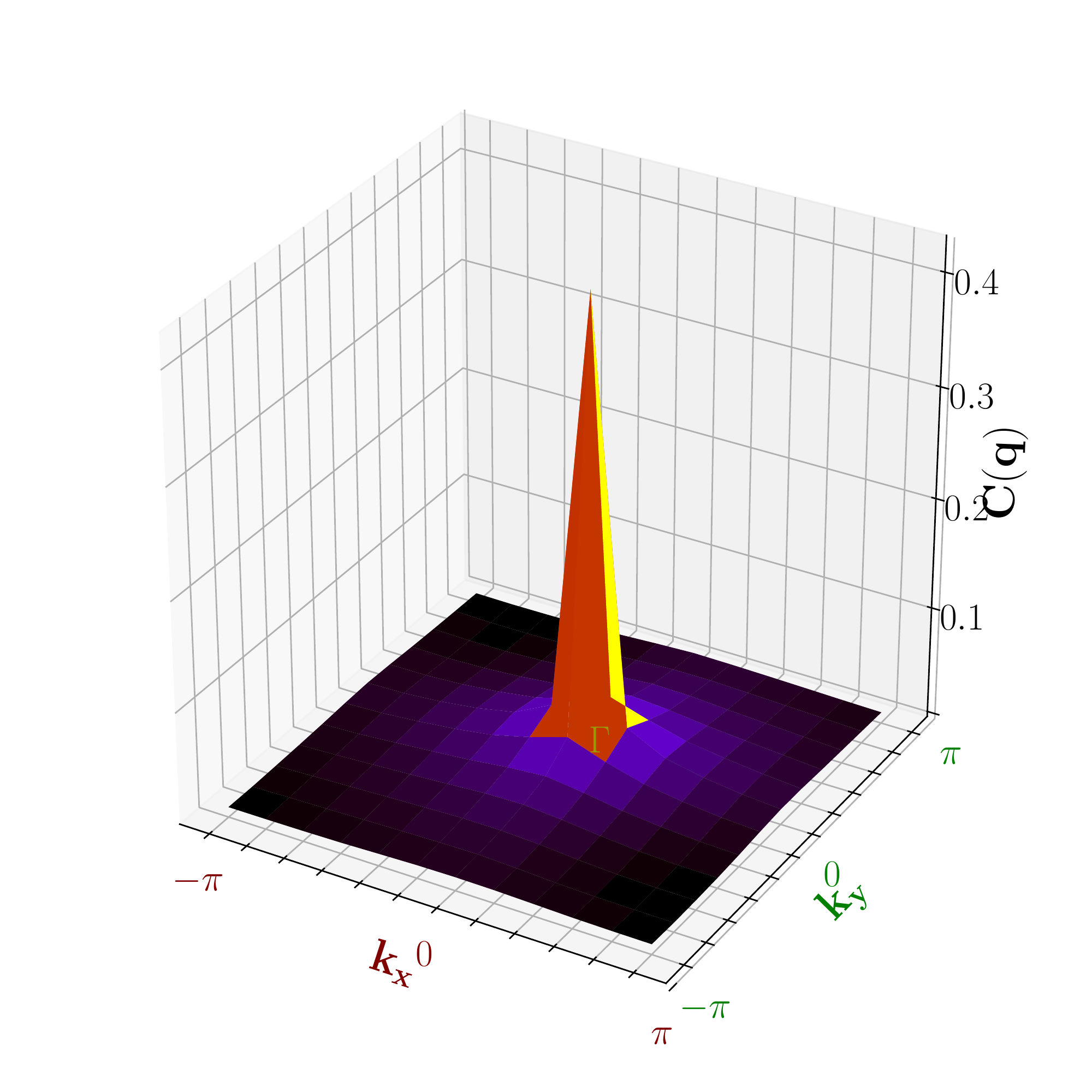}
  \caption{Fourier transform of $\tilde{C}(r)=\langle \psi(0) \psi(\vec{r})\rangle$ shows a Bragg peak at $\Gamma$ indicating
  breaking of rotational symmetry.}
 \end{figure}
 
 \begin{figure}
  \centering
  \includegraphics[scale=0.45]{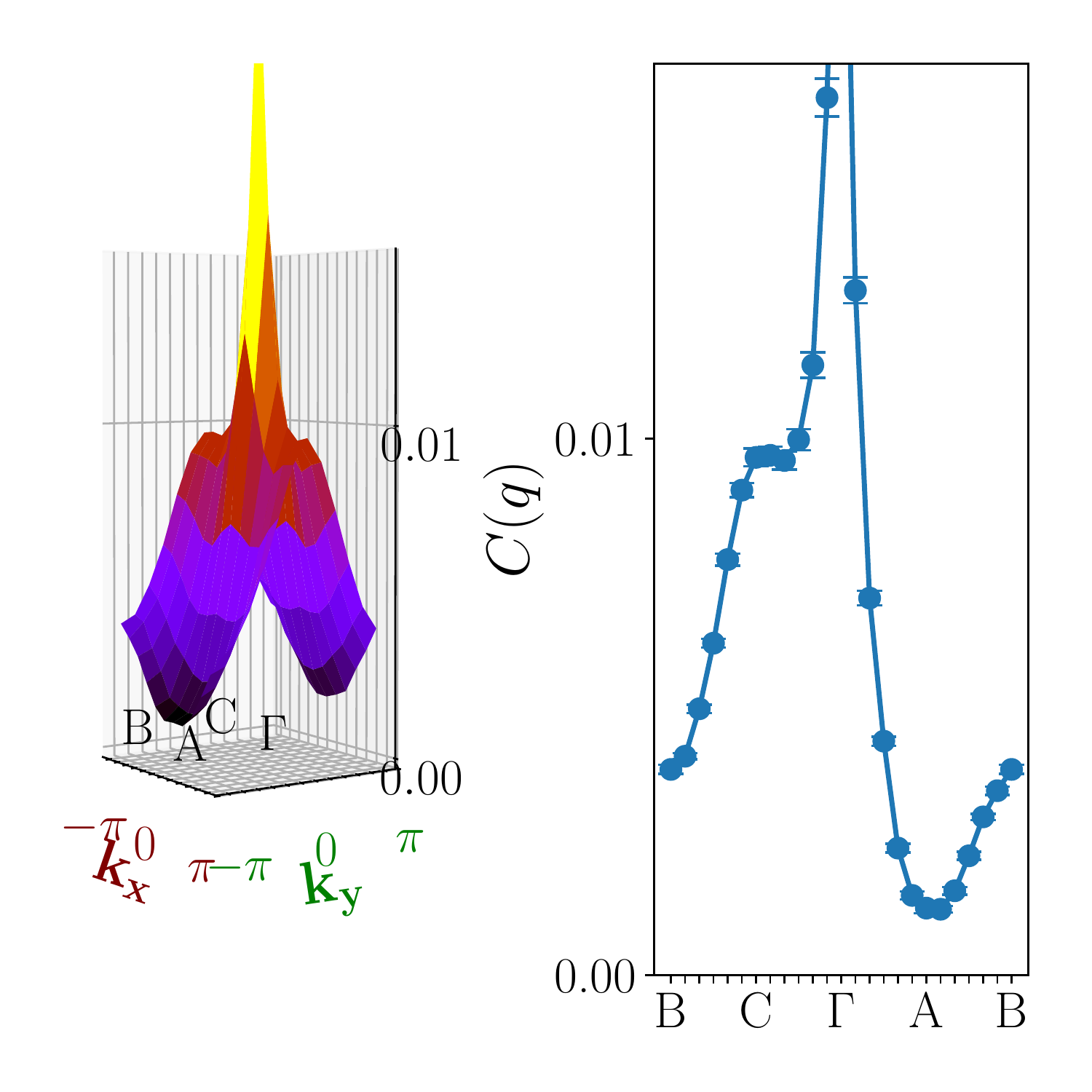}
  \caption{Fourier transform of $\tilde{C}(r)=\langle \phi(0) \phi(\vec{r})\rangle$ shows no peak except at $\Gamma$. We expect
  a Bragg peak at non-zero momentum if the state breaks translational
  symmetry. Therefore we conclude that our groundstate does break translational symmetry.}
 \end{figure}

\end{enumerate}

 \begin{table}
 \begin{tabular}{ c | c | c | c | c | c | c | c }
 \hline 
 Lattice & Lx & Ly & S & $e_{ED}$ &  $e_{QMC}$ & $\rho^{ED}_s$ & $\rho^{QMC}_s$   \\
 \hline
 Square & 4 & 4 & $\frac{1}{2}$ & -1.20178  & -1.20186(8) & 0.1855 & 0.1849(4)  \\
\hline
 Square & 2 & 2 & 1 & -5.0  & -5.0011(5) & 1.0 & 1.002(2)  \\
\hline
 Square & 2 & 2 & $\frac{3}{2}$ & -10.5  & -10.499(5) & 2.0 & 2.008(3)  \\
\hline
 Chain & 4 & 1 & 1 & -2.5  & -2.5004(2) & 0.2222 & 0.2226(4)    \\
\hline
 Chain & 6 & 1 & $\frac{3}{2}$  & -5.1488  & -5.1489(3) & 0.2630 & 0.2627(3)   
\end{tabular}
\caption{\label{tab1}Quantities measures by QMC at low temperatures (inverse temperature of $\beta=6 L$) for the Heisenberg Antiferromagnet with 
coupling constant $J=1$ compared with those determined for the ground state of the same model as found from Exact Diagonalization (ED).
The energies reported ($e_{ED}$ and $e_{QMC}$) are per site and the stiffness ($\rho^{ED}_s$ and $\rho^{QMC}_s$) are as described by Eq. \ref{winding}. 
The energy saturates to the ground state value for low enough temperatures as can be seen in Fig. \ref{finiteenergy}}
\end{table}

 \begin{figure}
 \includegraphics[scale=0.45]{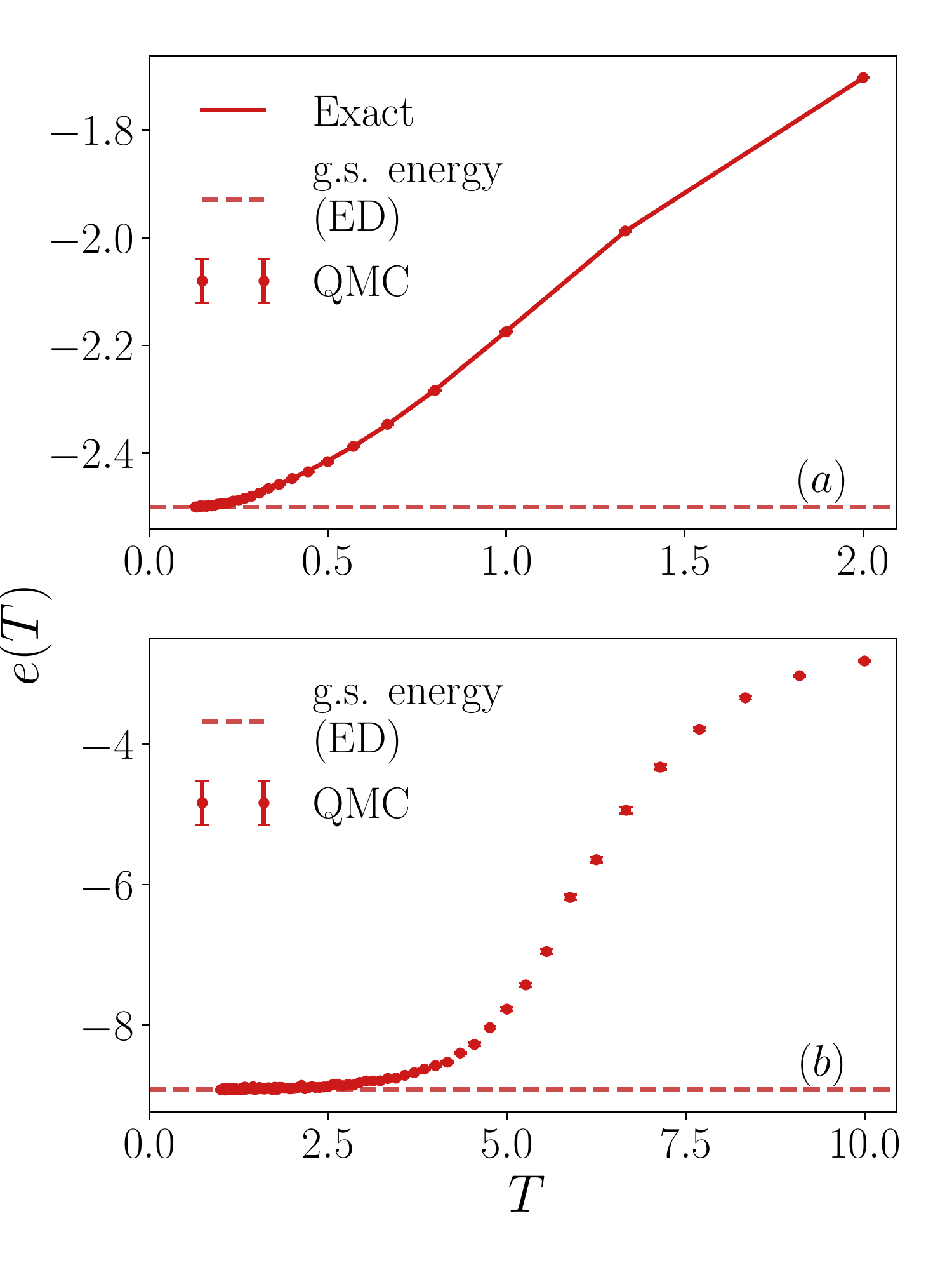}
 \caption{\label{finiteenergy}(a) Finite temperature energy (per unit site) comparison with ED for a 4 site spin-1 Heisenberg Antiferromagnetic chain
 (b) Ground state energy (per unit site) comparison for a 3$\times$3 square lattice with periodic boundary conditions (PBC) for the model decribed by a
 modified version of Eq. 6 of the main manuscript where $S_i.S_j$ replaced by $S_i^zS_j^z-\frac{1}{2}(S_i^{+}S_j^{+}+S_i^{-}S_j^{-})$ for $g=0.1$ (On a 
 bipartite lattice, this modification corresponds to a unitary transformation on one sublattice and hence simulating the modified model is no different from
 simulating the original model. However, since the $3\times3$ square lattice is non-bipartite, it is really the modified model that we simulate in QMC)}
\end{figure}

\begin{figure}
\centering
 \includegraphics[scale=0.35]{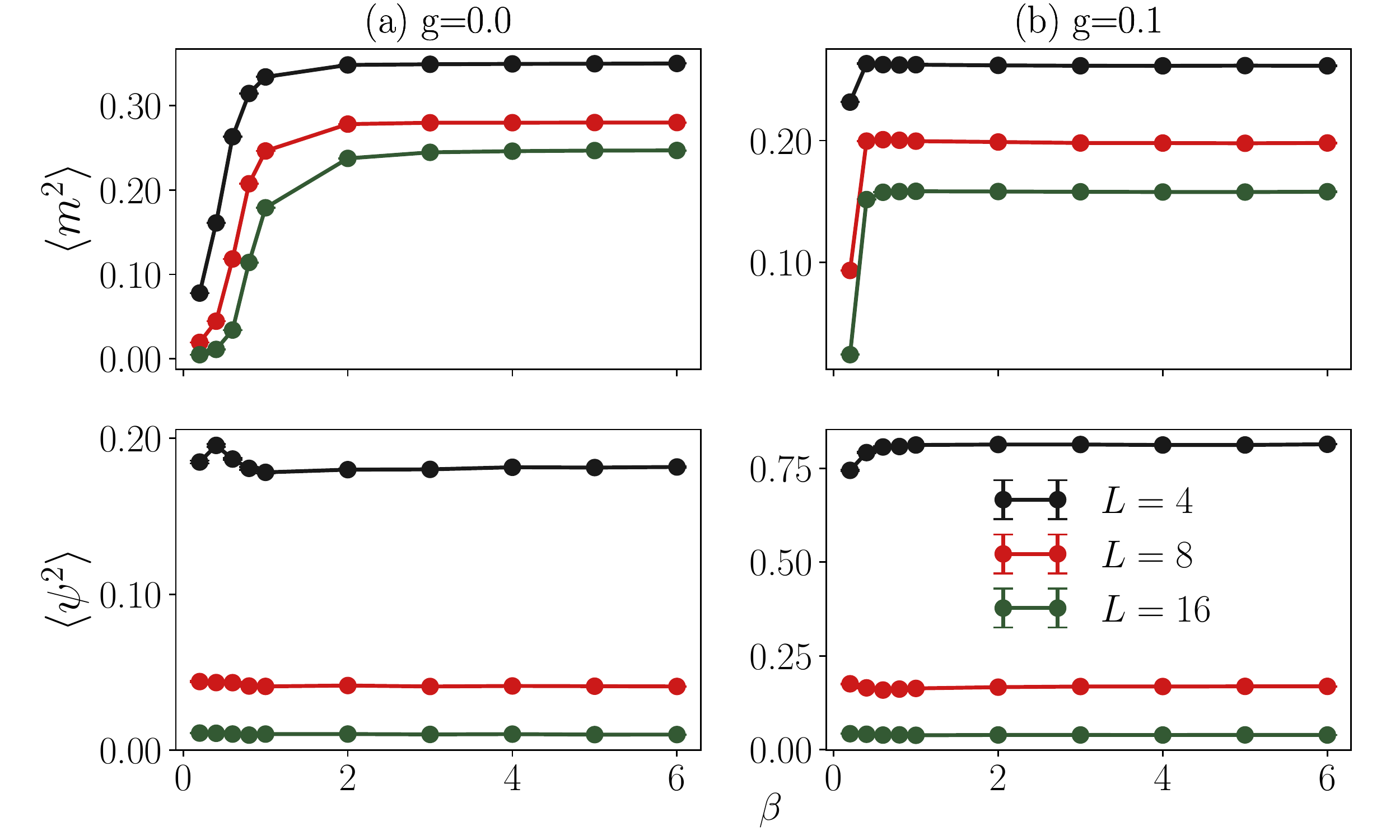}
 \caption{\label{magvsbeta1} Convergence of the order parameters ($\langle m^2 \rangle$ and $\langle \psi^2 \rangle$) as a function of inverse temperature for 
 (a) $g=0.0$ (b) $g=0.1$ : Finite temperature effects are clearly absent for $\beta \geq 4$, therefore we pick
 $\beta=4$ for our simulations to study behaviour in the zero temperature limit.}
\end{figure}

\begin{figure}
\centering
 \includegraphics[scale=0.4]{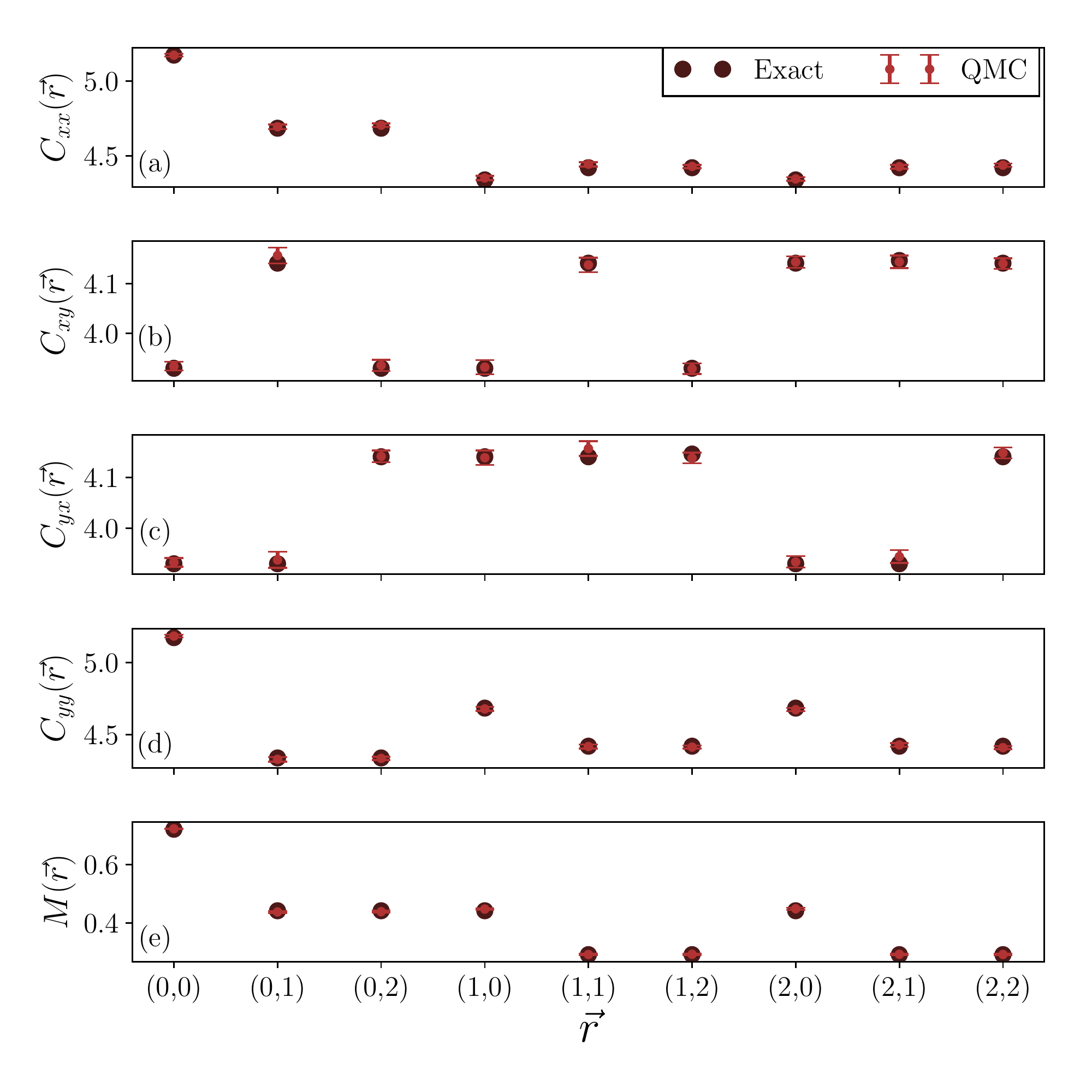}
 \caption{\label{square3by3corr}Correlation function comparison for a 3$\times$3 square lattice with periodic boundary conditions (PBC) for 
 a modified version of the model described by 
 Eq. 6 of the main manuscript for $g=0.1$ (the modification is as described in the caption of Fig. \ref{finiteenergy}):- 
 (a)-(d): $C_{ij}(\vec{r})=\langle B_i(0)B_j(\vec{r})\rangle$ where $B_i$ is as described in Eq. \ref{sing-proj}, (e) $M(r)$ is the 
 magnetic correlation function, $\langle S^z(\vec{0})S^z(\vec{r})\rangle$}
\end{figure} 

\subsection{Simulated Models}

\begin{figure}
 \includegraphics[scale=0.4]{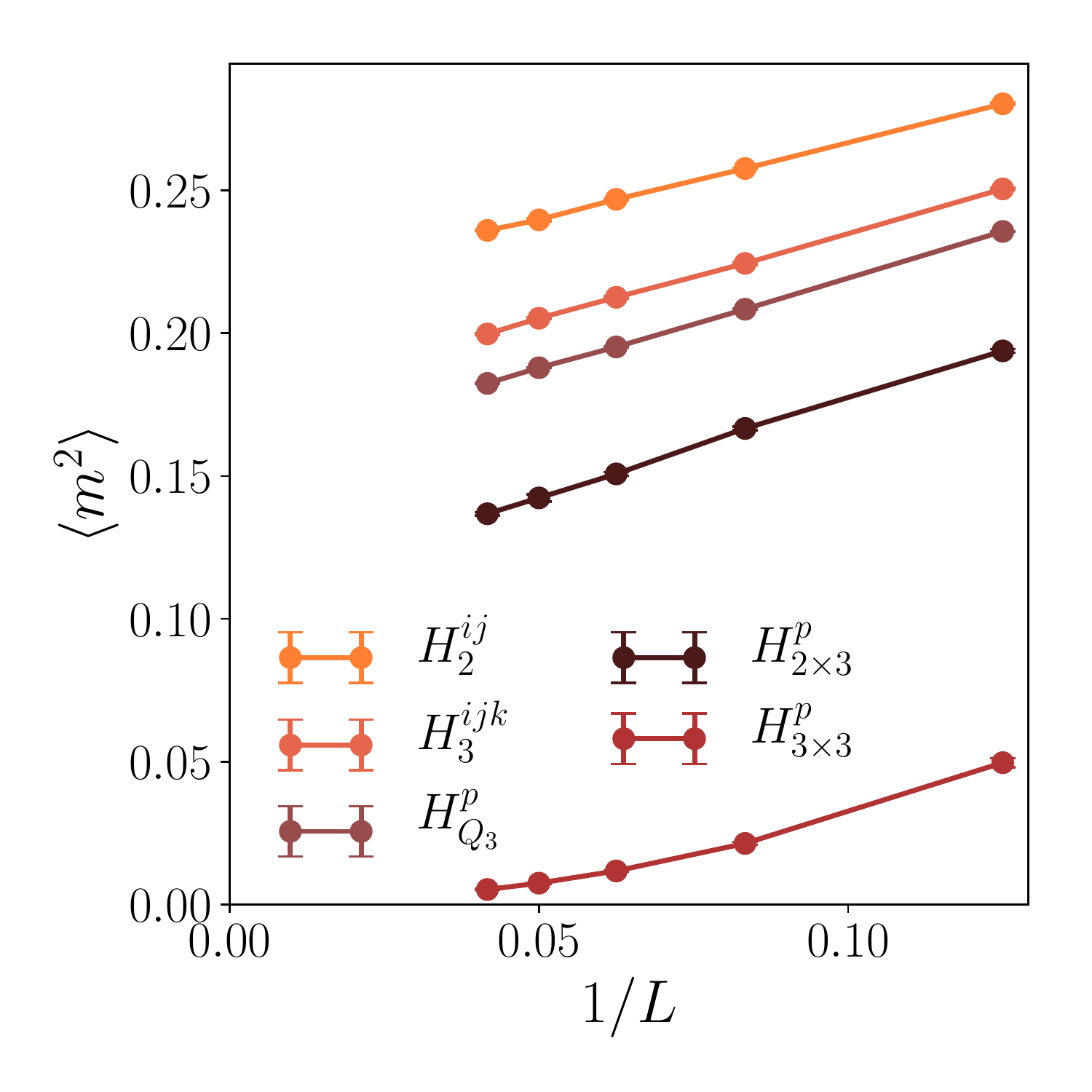}
 \caption{\label{magorder_allint}Magnetic order parameter extrapolation for $H^{ijk}_{3}$,$H^{p}_{2\times3}$,$H^p_{3\times3}$ and $H_{Q_3}$ compared
 with that for the Heisenberg model, $H^{ij}_2$: Only $H^p_{3\times3}$ is strong enough to destroy N\'eel order.}
\end{figure}

We simulated the $H^{ijk}_{3}$ interaction described by Eq. \ref{Hint2} and two other interactions constructed from it. One of them is the $3\times3$ plaquette
interaction described by $H^p_{3\times3}$ in Eq. 5 in the main manuscript. The other is a 6 spin plaquette version of the same interaction, 
$H^{p}_{2\times3}$, where a product of two $H^{ijk}_{3}$ is taken as shown in Eq. \ref{6spin}. 
We also study the spin-1 version of the $Q_3$ interaction (Eq. \ref{Q3}) introduced by Lou et. al. \cite{sandvik-kawashima}. 
In Eq. \ref{6spin} and Eq. \ref{Q3} the sites are numbered as in Fig. 1(e) of main manuscript. We see in Fig. \ref{magorder_allint}  that among all of these interactions only $H^p_{3\times3}$ successfully destroys the N\'{e}el order.

\begin{equation}
 H^p_{2\times3}= H^{123}_{3}H^{456}_{3}+H^{147}_{3}H^{258}_{3}
 \label{6spin}
\end{equation}

\begin{multline}
 H^{p}_{Q_3}= (1-\vec{S}_1.\vec{S}_2)(1-\vec{S}_4.\vec{S}_5)(1-\vec{S}_7.\vec{S}_8) \\
		  +(1-\vec{S}_1.\vec{S}_4)(1-\vec{S}_2.\vec{S}_5)(1-\vec{S}_3.\vec{S}_6)
 \label{Q3}
\end{multline}

\subsection{Order Parameter Collapses}

\begin{figure}[H]
 \includegraphics[scale=0.35]{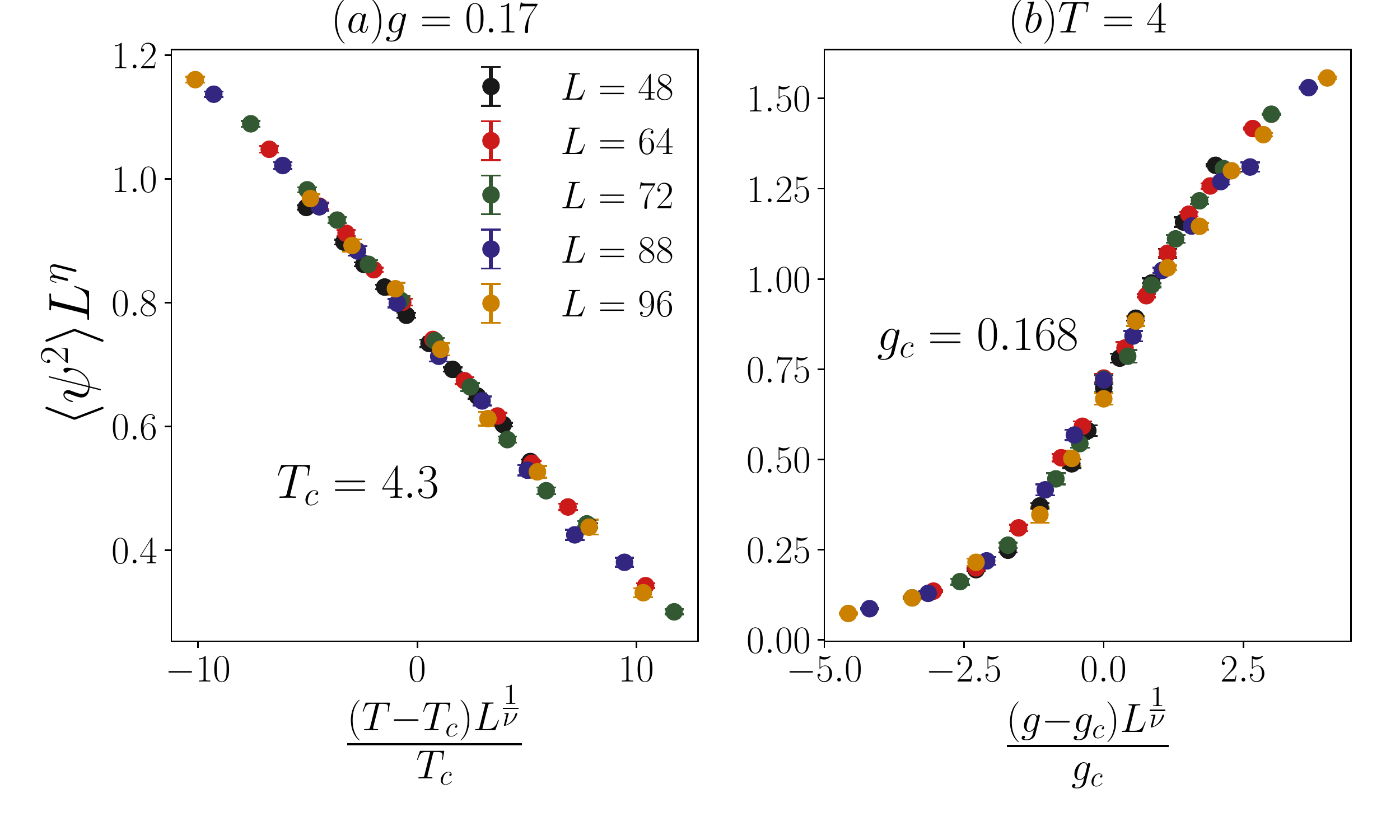}
 \caption{Order parameter scaling collapse with Ising critical exponents ($\nu=1$ and $\eta=\frac{1}{4}$) for transitions as a function of (a) temperature for a  fixed $g$ and (b) $g$ for a fixed temperature for the model described by Eq. 6 of main manuscipt}
\end{figure}

\putbib[supplementary.bib]
\end{bibunit}

\end{document}